\documentclass[11pt]{article}
\usepackage{mathrsfs}
\usepackage{amsmath}
\usepackage{amsfonts}
\usepackage{amssymb, amsmath, cite}

\newtheorem{theorem}{Theorem}

\newcommand\be{\begin{equation}}
\newcommand\ee{\end{equation}}
\newcommand\ber{\begin{eqnarray}}
\newcommand\eer{\end{eqnarray}}
\newcommand\berr{\begin{eqnarray*}}
\newcommand\eerr{\end{eqnarray*}}

 \newcommand\re{\mathrm{e}}
  \newcommand\ri{\mathrm{i}}
\newcommand{\ud}{\mathrm{d}}
\newcommand{\nm}{\nonumber}

\newcommand{\ito}{\int_{\Omega}}
\newcommand{\itr}{\int_{\mathbb{R}^2}}
 \newcommand\wot  {W^{1, 2}(\mathbb{R}^2)}
\newcommand{\vep}{\varepsilon}

\title{Existence Theorems for Vortices in\\the Aharony--Bergman--Jaferis--Maldacena Model}
\author{Xiaosen Han\\Institute of Contemporary Mathematics\\School of Mathematics\\Henan University\\Kaifeng, Henan 475004, PR China\\ \\Yisong Yang\\Department of Mathematics\\Polytechnic Institute of New York University\\Brooklyn,
New York 11201, USA }

\date{}

\begin{document}
\maketitle
\begin{quote}
{{{\bfseries Abstract.} A series of sharp existence and uniqueness theorems are established for the 
multiple vortex solutions 
in the supersymmetric Chern--Simons--Higgs theory formalism of Aharony, Bergman, Jaferis, and Maldacena, for which
 the Higgs bosons and Dirac fermions lie in the bifundamental representation of the general 
gauge symmetry group $U(N)\times U(N)$. The governing equations are of the BPS type and derived by
 Kim, Kim, Kwon, and Nakajima in the mass-deformed framework labeled by a continuous parameter.}}

\end{quote}

\section{Introduction}

It is well known that the presence of the Chern--Simons terms in field theories is essential in many areas of modern physics, especially
for condensed matter systems \cite{F,FM1,FM2,Wil}. More recently, the importance of the Chern--Simons terms 
in superstring theory and M-theory was explored in a general formalism by Schwarz \cite{Sch},
in the context of some Lagrangian descriptions of superconformal gauge field theories which couple
the Chern--Simons gauge and matter fields. This formalism was then made into fruition by Bagger and Lambert
\cite{BL1,BL2,BL3} and Gustavsson \cite{G}, which has since commonly been referred to as the Bagger--Lambert--Gustavsson (BLG) theory
\cite{BBKR,CCR,EMP}. Shortly afterwards, Aharony, Bergman,  Jafferis, and  Maldacena (ABJM) developed an elegant bi-level Chern--Simons--Higgs theory \cite{abjm}, extending the BLG theory. In both the BLG and ABJM theories, topological solitons realized
as the classical solutions of various BPS (named after the pioneering works of Bogomol'nyi \cite{Bo} and Prasad--Sommerfield \cite{PS}) equations
known as fuzzy funnels, domain walls, and vortices, are basic field configurations that describe M-branes. 
In the present study, we aim at establishing an existence and uniqueness theory for the multiple vortex solutions in a general
class of BPS equations in the ABJM model.
It is interesting to note that, in these superconformal Lagrangian field theories \cite{abjm,BL1,BL2,BL3,Sch}, the dynamics of the gauge fields is
exclusively of the Chern--Simons type, meaning that the usual quadratic kinetic terms involving the gauge field strength
tensors, are missing \cite{Sch}, as in the earlier Chern--Simons--Higgs models of Hong--Kim--Pac \cite{HKP} and
Jackiw--Weinberg \cite{JW}. In these latter studies \cite{HKP,JW} and the subsequent development of the subject
(cf. \cite{Dunne} and references therein), suitable six-order Higgs potential density functions 
have to be taken in order to achieve a BPS reduction. While, in such a situation, the BPS reductions can be made, the six-order
potential terms lead to several complications or drawbacks of the models. The first one is that it is not clear whether
topological solutions are uniquely determined by any prescribed distribution of vortices, as seen in the classical Abelian
Higgs theory \cite{jata,T1}, although the existence of maximal solutions has been established \cite{SYcs1}.
The second one is the occurrence of non-topological solutions \cite{JPW,SYcs2} which are plagued by
non-uniqueness and the issue of existence of solutions realizing arbitrarily prescribed vortices has only been
tackled in some extreme cases \cite{Chae,CFL}. The third one concerns the solutions over a doubly-periodic domain
resembling a vortex condensate/lattice structure \cite{Ab,CY,waya} and it is known that 
even in such a compact setting non-uniqueness occurs \cite{Ta} and the interval ensuring the existence of a solution
for the Chern--Simons coupling constant cannot be explicitly determined \cite{CY,Ta}. Furthermore, 
for non-Abelian BPS vortex equations, although the existence
of topological solutions can be proved \cite{Ycs}, the existence of vortex condensates has only been established for
rank 2 gauge groups \cite{NT} and the existence of non-topological solutions is still unsettled \cite{Tbook,yang1}.
For the Chern--Simons vortices arising in the ABJM theory, however, we shall see through
the present work that the situation is totally
different. More precisely, we will develop a complete existence theory for multiple vortex solutions arising in
the ABJM theory with the general gauge group $U(N)\times U(N)$ which gives us a  unique topological solution for
any prescribed distribution of vortices in the plane and a unique doubly periodic condensate solution under explicitly
stated necessary and sufficient conditions involving several related physical coupling parameters. Furthermore,
the model will be seen to be free of non-topological
solutions.

The governing BPS equations considered here for the multiple vortex solutions in the ABJM theory were discovered in 
the work of Kim--Kim--Kwon--Nakajima \cite{kkkn} who showed that for the bottom case when the gauge group is $U(2)\times U(2)$
the equations can be reduced to that of the Abelian Higgs equation \cite{GST,jata,NO} which has been thoroughly understood \cite{jata,T1}
and that for the general case with the gauge group $U(N)\times U(N)$ (with $N\geq3$) the equations are given as an
$(N-1)$ by $(N-1)$ system
of nonlinear elliptic equations of delicate structures. These equations arise in the so-called `mass-deformed' theory labeled by
a mass deformation parameter $\mu\geq0$. When $\mu=0$ (mass deformation is absent), it has been shown in\cite{kkkn} that
there is no finite-energy regular solution, and that finite-energy solutions can only be expected to appear when
$\mu>0$ (mass deformation is present). The aim of our work here
is to establish an existence theory for such mass-deformed ABJM vortex equations.
A similar structure has also been seen in the independent work of Auzzi and
Kumar \cite{ak} for which some existence and uniqueness theorems have been obtained in \cite{liey1}
through exploiting appropriate variational methods. Enlightened by the success of that study, our technical strategy in the current study will
again be
to explore and unveil the underlying variational structures of the system. It is interesting to notice that, although
the Higgs potential density function still contains six-order terms, which play  a crucial role for generating non-topological vortices in the classical Chern--Simons--Higgs models \cite{JPW,SYcs2}, such vortices are absent
in the ABJM situation.

The content of the rest of the paper is outlined as follows. In Section 2, we recall the multiple vortex equations derived in the work
of Kim--Kim--Kwon--Nakajima \cite{kkkn} in the ABJM model \cite{abjm} and state our main existence results. The subsequent
sections are then devoted to proofs of results. Specifically, in Section 3, we first apply a matrix decomposition
procedure to unveil the variational structure of the system of governing elliptic equations. 
We next prove the existence of a solution by a somewhat indirect minimization approach,
using the method of \cite{jata,T1}. We then obtain the asymptotic
decay properties of the solution on the full plane and calculate the anticipated quantized integrals which give rise
to quantized vortex fluxes. In Section 4, we consider the compact case when solutions are doubly periodic. We shall
mainly adapt
the direct minimization method used in \cite{liey1}. In Section 5, we present the limiting case $a=0$ which is of 
independent interest. We will see that such a limiting case allows us to state our necessary and sufficient conditions
for the existence of a doubly periodic multiple solution explicitly.
In Section 6, we reconsider
the case when $a>0$ and present two concrete examples, with $N=3$ and $N=4$, respectively, as an illustration of the
application of our results in the general situation ($a>0$). We shall also describe how to use our results to estimate
the dimension of the moduli space of the BPS equations in the ABJM model under consideration.

\section{Existence of vortices in the ABJM model}
\setcounter{equation}{0}\setcounter{theorem}{0}

Use $\mu,\nu=0,1,2$ to denote the Lorentzian indices of the Minkowski spacetime $\mathbb{R}^{2,1}$ of signature
$(-++)$. Like the BLG model \cite{BL1,BL2,BL3,G}, the ABJM model  \cite{abjm,BLS} is formulated as a low-energy approximation of
  multiple M2-branes so that it is dual to M-theory on appropriate anti-de Sitter orbifolds.  It is an $\mathcal{N}=6$ supersymmetric Chern--Simons theory with the gauge group $U(N)\times U(N)$ governed by the full Lagrangian density
\ber
   \mathcal{L}_{\mbox{ABJM}}=\mathcal{L}_{\mbox{CS}}+\mathcal{L}_{\mbox{kin}}-V_{\mbox{ferm}}-V_0,\label{e1}
   \eer
in which the Chern--Simons Lagrangian ${\cal L}_{\mbox{CS}}$ is given by
   \ber
   \mathcal{L}_{\mbox{CS}}= \frac{k}{4\pi}\epsilon^{\mu\nu\lambda}\mathrm{Tr}\left(A_\mu\partial_\nu A_\lambda+\frac{2\ri}{3}A_\mu A_\nu A_\lambda
   -\hat{A}_\mu\partial_\nu \hat{A}_\lambda-\frac{2\ri}{3}\hat{A}_\mu \hat{A}_\nu \hat{A}_\lambda \right), \label{e2}
   \eer
which describes two Lie algebra $u(N)$-valued
gauge fields, $A_\mu$ and $\hat{A}_\mu$, with the
Chern--Simons level $(k,-k)$; using $\gamma^\mu$ to denote the 
Dirac matrices expressible in terms of the standard Pauli spin matrices,
$\gamma^0=\ri\sigma^2,\gamma^1=\sigma^1,\gamma^2=\sigma^3$, the matter-kinetic Lagrangian ${\cal L}_{\mbox{kin}}$ is given by
    \ber
    \mathcal{L}_{\mbox{kin}}=-\mathrm{Tr}\left(D_\mu Y_A^\dagger D^\mu Y^A\right) +\ri \mathrm{Tr}\left(\psi^{A\dagger}\gamma^\mu D_\mu\psi_A\right),\label{e3}
    \eer
which
couple four complex scalars  $Y_A$ $(A=1, 2, 3, 4)$ and four
Dirac fermions $\psi_A$ $(A=1, 2, 3, 4)$ in the bifundamental representations $(\mathbf{N}, \overline{\mathbf{N}})$
of the gauge group so that the gauge-covariant derivatives take the form
\be
 \left.\begin{array}{rrl} D_\mu Y^A &=&\partial_\mu Y^A+\ri A_\mu Y^A-\ri Y^A\hat{A}_\mu,\\
D_\mu Y_A &=&\partial_\mu Y^A+\ri \hat{A}_\mu Y_A-\ri Y_A{A}_\mu,  \end{array}\right\}\quad A=1,2,3,4;\label{e4}
  \ee
the Yukawa-like quartic-interaction potential density $V_{\mbox{ferm}}$   is given by
  \ber
  V_{\mbox{ferm}}&=& \frac{2\pi \ri}{k}\mathrm{Tr}\left( Y_A^\dagger Y^A\psi^{B\dagger}\psi_B-Y^AY_A^\dagger \psi_B\psi^{B\dagger}
     +2Y^AY_B^\dagger\psi_A\psi^{B\dagger} \right.\nm\\
      &&\left. -2Y_A^\dagger Y^B\psi^{A\dagger}\psi_B -\epsilon^{ABCD}Y^\dagger_A \psi_BY^\dagger_C\psi_D+\epsilon_{ABCD}Y^A\psi^{B\dagger}Y^C\psi^{D \dagger}\right);\label{e5}
   \eer
 and $V_0$ is the sextic scalar potential
  \ber
   &&V_0=\nm\\
&&\frac{4\pi^2}{3k^2}\mathrm{Tr}\left(6Y^AY_B^\dagger Y^B Y^\dagger_A Y^CY^\dagger_C -Y^AY_A^\dagger Y^BY_B^\dagger Y^CY_C^\dagger -Y_A^\dagger Y^AY_B^\dagger Y^BY_C^\dagger Y^C  -4Y^AY_B^\dagger Y^C Y_A^\dagger Y^BY_C^\dagger \right).\nm \\
\label{e6}
   \eer

With the Lagrangian density  given by (\ref{e1}), the action 
\ber 
{\cal A}_{\mbox{ABJM}}=\int{\cal L}_{\mbox{ABJM}}\,\ud x,
\eer
evaluated over the full spacetime
is invariant under the  $\mathcal {N}=6$ supersymmetry transformation \cite{abjm,kkkn}
  \ber\left.\begin{array}{lll}
  \delta Y^A&=&\ri \omega^{AB}\psi_B,\\
  \delta\psi_A&=&-\gamma_\mu\omega_{AB}D_\mu Y^B+\frac{2\pi}{k}\left(-\omega_{AB}\left[Y^CY_C^\dagger Y^B-Y^BY^\dagger_CY^C\right]+2\omega_{BC}Y^BY^\dagger_AY^C\right)\\
  &\equiv&-\gamma_\mu\omega_{AB}D_\mu  Y^B+\omega_{BC}\left(\beta_A^{BC}+\delta_A^{[B}\beta_D^{C]D}\right),\\
  \delta A_\mu&=&-\frac{2\pi}{k}\left(Y^A\psi^{B\dagger}\gamma_\mu\omega_{AB}+\omega^{AB}\gamma_\mu\psi_AY^\dagger_B\right),\\
  \delta \hat{A}_\mu&=&\frac{2\pi}{k}\left(\psi^{A\dagger}Y^B\gamma_\mu\omega_{AB}+\omega^{AB}\gamma_\mu Y^\dagger_A\psi_B\right),
 \end{array}\right\}\label{e8}
  \eer
 where $\omega_{AB}$ are supersymmetry transformation parameters  satisfying
  \be
   \omega^{AB}=(\omega_{AB})^*=-\frac12\epsilon^{ABCD}\omega_{CD},\label{e9}
  \ee
and $\beta_C^{AB}$ are given by the expressions
 \be
  \beta_C^{AB}=\frac{4\pi}{k}Y^{[A}Y^\dagger_CY^{B]}=\frac{4\pi}{k}\left(Y^AY^\dagger_C Y^B-Y^BY^\dagger_CY^A\right).\label{e10}
 \ee

In the mass-deformed theory, we need to update the Lagrangian density (\ref{e1})
by modifying the potential densities $V_{\mbox{ferm}}$ and $V_0$  following the recipe
\ber
V_{\mbox{ferm}}\mapsto V_{\mbox{ferm}}+\Delta V_{\mbox{ferm}},\quad V_0\mapsto V_0+\Delta V_0\equiv V_m,
\eer
where
\ber
  \Delta V_{\mbox{ferm}}&=&\mathrm{Tr}\left(\mu\psi^{\dagger A}M_A^B\psi_B\right),\label{e12}\\
  \Delta V_0&=& \mathrm{Tr}\left(\frac{4\pi \mu}{k}Y^AY^\dagger_A Y^BM_B^CY^\dagger_C
  -\frac{4\pi \mu}{k}Y^\dagger_A Y^A Y_B^\dagger M^B_CY^C+\mu^2Y^\dagger_AY^A\right),\label{e13}
 \eer
in which $\mu>0$ is a mass-deformation parameter which should not be confused
with the Lorentzian index and $M_A^B=\mathrm{diag}\{1, 1, -1, -1\}$. We notice that, although the original
Higgs potential density $V_0$ given in (\ref{e6}) is purely sextic, the mass-deformed potential density 
$V_m$ obtained from adding the correction term $\Delta V_0$ given in (\ref{e13}) contains both quadratic and quartic
terms, as that in the classical Chern--Simons--Higgs model \cite{HKP,JW}.

 It is evident that the associated Euler--Lagrange
equations of the mass-deformed action 
\be 
{\cal A}_{\mbox{mass-deformed ABJM}}=\int\left\{{\cal L}_{\mbox{CS}}+{\cal L}_{\mbox{kin}}-(V_{\mbox{ferm}}+\Delta V_{\mbox{ferm}})-V_m\right\}\,\ud x,
\ee
are of course rather complicated. In the work of 
  Kim, Kim, Kwon, and Nakajima \cite{kkkn}, it is shown that these equations in their static limit and
in the absence of fermions may
be reduced into the following remarkable BPS system of equations
  \ber\left.\begin{array}{rrl}
   (D_1-\ri s D_2)Y^1&=&0, \\
   D_iY^A&=&0,\quad (A\neq1, i=1,2),\\
   D_0Y^1+\ri s\left(\beta_2^{21}+\mu Y^1\right)&=&0, \\
\quad  D_0Y^2-\ri s\left(\beta_1^{12}+\mu Y^2\right)&=&0, \\
   D_0Y^3-\ri s \beta_1^{13}&=&0,\\ D_0Y^4-\ri s \beta_1^{14}&=&0,\\
   \beta_3^{31}&=&\beta_4^{41}=\beta_3^{21}+\mu Y^1,\\
\quad \beta_4^{43}&=&\mu Y^3,\\
 \quad \beta_3^{34}&=&\mu Y^4, \\
   \beta_3^{32}=\beta_4^{42}=\beta_2^{23}=\beta_2^{24}&=&0,\\
 \beta_A^{BC}&=&0, \quad (A\neq B\neq C\neq A)\end{array}\right\}\label{BPS}
  \eer
 where $s=\pm1$ is a signature symbol to be specified later, coupled with the usual Gauss law constraints
 \ber
  \frac{\kappa}{2\pi}B\equiv\frac{\kappa}{2\pi}F_{12}&=&j^0, \label{e20}\\  \frac{\kappa}{2\pi}\hat{B}\equiv \frac{\kappa}{2\pi}\hat{F}_{12}&=&-\hat{j}^0,\label{e21}
 \eer
where $j^0$ and $\hat{j}^0$ are two associated  matrix-valued conserved currents given by the expressions
 \ber
  j_\mu=\ri \left(Y^AD_\mu Y^\dagger_A-D_\mu Y^AY_A^\dagger\right),  \label{e22}\\
 \hat{j}_\mu=\ri \left(Y_A^\dagger D_\mu Y^A-D_\mu Y_A^\dagger Y^A\right).\label{e23}
 \eer

In \cite{kkkn}, it is demonstrated that the energy of the mass-deformed ABJM model has the topological lower bound
\be \label{low}
E\geq\frac13\mu|Q+2R_{12}|,
\ee
where the topological charges $Q$ and $R_{12}$ are given by
\ber 
Q&=&\mbox{Tr}\int j^0\,\ud x,\\
R_{12}&=&\mbox{Tr}\int J^0_{12}\,\ud x,
\eer
for which the charge densities $j^0$ is defined in (\ref{e22}) and $J^0_{12}$ given by
\be 
J^0_{12}=\ri (Y^1D_0Y^\dagger_1-D_0 Y^1 Y^\dagger_1)-\ri(Y^2D_0Y^\dagger_2-D_0 Y^2 Y^\dagger_2),
\ee
and the integration is carried over the full two-dimensional spatial domain, and the lower bound (\ref{low}) is attained
by the solutions of the BPS system (\ref{BPS}) coupled with the Gauss law constraints (\ref{e20})--(\ref{e21}) so that
$s$ is determined by the condition
\be 
|Q+2R_{12}|=s(Q+2R_{12}).
\ee

Thus it is important to understand the solutions of (\ref{BPS})--(\ref{e21}) which will be our goal in the
present work.

To approach the system of equations (\ref{BPS})--(\ref{e21}), Kim, Kim, Kwon, and Nakajima \cite{kkkn}  take the 
following ansatz to represent the $N\times N$ matrices $Y^A=(Y^A_{ij})$:
  \ber
  Y^1_{ij}=\delta_{i+1,j}\sqrt{\frac{k\mu}{2\pi}}f_i, \quad Y^2_{ij}=\delta_{ij}\sqrt{\frac{k\mu}{2\pi}}a_i,\quad Y^3=0,\quad Y^4=0, \label{e24}
  \eer
where  $f_i$  ($i=1, \dots, N-1$)  are complex-valued  functions and
 \be
  a_i=\sqrt{a^2+i-1}, \quad i=1, \dots, N,\label{e25}
 \ee
 with $a\ge 0$ a constant.
Within this ansatz,   the $N\times N$ matrix-valued `magnetic' fields become diagonal whose entries are given by
\cite{kkkn}
   \be
    B_{ij}=\hat{B}_{ij}=-2s\delta_{ij}\mu^2(a^2+i-1)(|f_i|^2-|f_{i-1}|^2+1),\quad  i, j=1,\dots,N,\label{e26}
   \ee
 where  the convention $f_0=f_N=0$ is imposed.  Using the complex-variable differentiation
  $
   \partial=\frac12(\partial_1-\ri \partial_2), \overline{\partial}=\frac12(\partial_1+\ri \partial_2)
  $, and eliminating the gauge fields from the equations,  it is shown in \cite{kkkn} that
the BPS system (\ref{BPS})--(\ref{e21}) is reduced into the following system of    $N-1$ ($N\geq3$) coupled  vortex  equations:
  \ber
  \partial\bar{\partial}\ln|f_1|^2&=&\mu^2\left(\left[2a^2+1\right]|f_1|^2-\left[a^2+1\right]|f_2|^2-1\right), \label{e28}\\
  \partial\bar{\partial}\ln|f_i|^2&=&\mu^2\left(-\left[a^2+i-1\right]|f_{i-1}|^2+\left[2a^2+2i-1\right]|f_i|^2-\left[a^2+i\right]|f_{i+1}|^2-1\right),\nm\\
  &&i=2, \dots, N-2, \label{e29}\\
  \partial\bar{\partial}\ln|f_{N-1}|^2&=&\mu^2\left(-\left[a^2+N-2\right]|f_{N-2}|^2+\left[2a^2+2N-3\right]|f_{N-1}|^2-1\right)\label{e30}
  \eer
  away from the zero points of $f_1,  \dots, f_{N-1}$, which are known to be the vortex points of the system.

 Analyzing the structure of the system (\ref{BPS}) as in \cite{jata},
it may be seen that  the zeros of the fields $f_1,\dots, f_{N-1}$ are discrete and of
 integer multiplicities. Thus we can denote the sets of zeros of  each $f_i$ by
 \ber
  Z_{f_i}=\big\{p_{i,1}, \dots, p_{i, n_i}\big\}, \quad i=1, \dots, N-1, \label{e31}
 \eer
such that the number of repetitions of any point $p$ among the set $Z_{f_i}=\{p_{i,s}\}$ ($i=1, \dots, N-1)$  takes account of the multiplicities of the zero.

We aim to prove that the prescribed sets of zeros given by \eqref{e31}
completely characterize the solution of \eqref{e28}--\eqref{e30}. To be precise
we note that the problem may be considered either over the full plane ${\mathbb{R}}^2$ under
the natural boundary condition
\be \label{BB}
\lim_{|x|\to\infty}(\partial\overline{\partial}\ln |f_i|^2)(x)=0,\quad i=1,\dots,N-1,
\ee
 or over a doubly periodic domain $\Omega$ so that the field configurations are subject to the 't Hooft periodic boundary
condition \cite{hoof,waya,yang1} for which  periodicity is achieved modulo gauge transformations. We shall establish
that in both cases solutions exist and are unique.

Note that (\ref{e28})--(\ref{e30}) only make sense when $N\geq3$. When $N=2$ (the `bottom' case
with the gauge group $U(2)\times U(2)$), the system is a single
equation \cite{kkkn}:
\be 
\partial\overline{\partial}\ln|f|^2=\mu^2([2a^2+1]|f|^2-1),
\ee
which has been well studied and existence and uniqueness results have been obtained
 \cite{jata,T1,waya}.

We now proceed to state our main results. 

Let $R$ be the $(N-1)\times(N-1)$ tridiagonal matrix given by
\ber
&&R=\nm\\
&&\begin{pmatrix}
2a^2+1 & -(a^2+1) & 0&\dots&\dots & 0\\
-(a^2+1) & 2a^2+3 & -(a^2+2)&0&\dots & 0\\
0 &-(a^2+2)&  2a^2+5 &-(a^2+3)&\dots&0\\
\vdots& &\ddots&\ddots&\ddots&\vdots\\
0&\quad &\ddots&-(a^2+N-3)&2a^2+2N-5&-(a^2+N-2)\\
 0 & \dots &   &0&-(a^2+N-2)&2a^2+2N-3
\end{pmatrix}.\nm\\
\eer
We  will see later that  the  matrix $R$ is  positive definite. Denote the inverse of $R$ by
$R^{-1}$. We shall also see that all entries  of $R^{-1}$ are  positive, i.e., $(R^{-1})_{ij}>0, \, i,j=1,\dots, N-1$.  Write the eigenvalues of
$R$ as $\lambda_1, \dots, \lambda_{N-1}$, and set $\lambda_0$ to be the positive quantity
 \be
 \lambda_0=2\min\{\lambda_1,\dots, \lambda_{N-1}\}.\label{e32}
 \ee

 Our main results are collectively summarized as follows.

  \begin{theorem}\label{tha1}
  For any $a\ge0, \mu>0$, consider  the system of multiple vortex equations \eqref{e28}--\eqref{e30} for  the field configuration $(f_1, \dots, f_{N-1})$ with the prescribed
  zero sets given  by  \eqref{e31} such that each $f_i$ has $n_i$ arbitrarily distributed zeros
$p_{i,1},\dots,p_{i,n_i}$, $i=1, \dots, N-1$.

  (i) For the problem  over the full plane $\mathbb{R}^2$,  there exists a  unique solution satisfying the boundary condition
  \be
   |f_i|^2\to r_i, \quad |x|\to \infty, \quad i=1,\dots,N-1,\label{e33}
  \ee
   which realizes the boundary condition (\ref{BB}), where
    \be \label{e33a}
r_i\equiv\sum\limits_{j=1}^{N-1}(R^{-1})_{ij}>0, \quad i=1, \dots, N-1.\ee
 Moreover, this  boundary condition is achieved exponentially fast at infinity,
    \ber
    \left||f_i|^2-r_i \right|\le  C(\vep)\re^{-2\mu(1-\vep)\sqrt{\lambda_0}|x|}\quad\mbox{as }|x|\to\infty, \quad i=1, \dots, N-1, \label{e34}
     \eer
 where $\vep\in (0,1)$ is arbitrarily small, $C(\vep)$ is a positive constant depending on $\vep$, and $\lambda_0$ is  defined by \eqref{e32}.

  (ii)  Over a doubly periodic  domain $\Omega$, the problem   admits  a unique  solution if and only if
 \ber
   \pi\sum\limits_{j=1}^{N-1}(R^{-1})_{ij}n_j<\mu^2|\Omega|\sum\limits_{j=1}^{N-1}(R^{-1})_{ij}, \quad i=1, \dots, N-1, \label{e35}
 \eer
hold simultaneously.

  (iii) In both cases, there hold the quantized integrals
  \be
   \int\left(\sum\limits_{j=1}^{N-1}R_{ij}|f_j|^2-1\right)\ud x=-\frac{\pi n_i}{\mu^2}, \quad i=1, \dots, N-1, \label{e36}
  \ee
where the integration is evaluated either over the full plane $\mathbb{R}^2$ or the doubly periodic domain $\Omega$.
 \end{theorem}

It can be checked that (\ref{e33}) and (\ref{BB}) are equivalent. Thus, by virtue of (\ref{e33a}), we see that the
so-called non-topological solutions \cite{Chae,CFL,Dunne,JPW,SYcs2} do not appear in the ABJM model \cite{abjm,kkkn}
considered here.

We now proceed to compute the associated flux.

As noticed in \cite{kkkn}, the ansatz taken makes it consistent
to assume that the gauge fields $A_l$ and $\hat{A}_l$ ($l=1,2$) are diagonal:
\be 
A_l=\hat{A}_l=\mbox{diag}\{a_l^1,\dots,a_l^N\},\quad l=1,2.
\ee
Thus, we may introduce the complex-valued variable
\be 
A=\frac12(A_1-\ri A_2)=\mbox{diag}\left\{\frac12(a_1^1-\ri a_2^1),\dots,\frac12(a_1^N-\ri a_2^N)\right\}\equiv
\{b_1,\dots,b_N\},
\ee
so that the matrix-valued `magnetic' field $B=F_{12}$ becomes
\be \label{Bb}
B=F_{12}=\partial_1 A_2-\partial_2 A_1=-2\ri (\partial\overline{A}-\overline{\partial}A)=-2\ri\mbox{ diag}\{
\partial \overline{b}_1-\overline{\partial}b_1,\dots,\partial\overline{b}_N-\overline{\partial}b_N\},
\ee
which should not be confused with the group index used earlier. Hence, in view of (\ref{BPS}), we have
\be \label{fi}
\ri\overline{\partial} \ln f_{i-1}=(\overline{b}_{i-1}- \overline{b}_i),\quad i=2,\dots,N,
\ee
away from the zeros of $f_i$ ($i=2,\dots,N$), where we have chosen $s=-1$ for definiteness. From (\ref{fi}), we obtain
\be\label{dfi}
\frac14\ri\,\Delta \ln|f_{i-1}|^2=\ri\partial\overline{\partial}(\ln f_{i-1}+\ln\overline{f}_{i-1})
=(\partial\overline{b}_{i-1}-\overline{\partial}b_{i-1})-(\partial\overline{b}_i-\overline{\partial}b_i),\quad i=2,\dots,N,
\ee
again away from the zeros of the functions. Comparing (\ref{Bb}) and (\ref{dfi}), we see that $B\equiv\mbox{diag}\{B_1,\dots,
B_N\}$ over the same domain can be expressed as
\be 
B_i=B_{i-1}-\frac12\Delta\ln|f_{i-1}|^2=B_1-\frac12\sum_{j=1}^{i-1}\Delta(\ln|f_j|^2),\quad i=2,\dots,N,
\ee
where $B_1$ may be read off from (\ref{e26}). That is, $B_1=2a^2\mu^2(|f_1|^2+1)$. Consequently, we get
\be 
\mbox{Tr}(B)=\sum_{i=1}^N B_i=NB_1-\frac12\sum_{i=1}^{N-1}(N-i)\Delta\ln|f_i|^2.
\ee

  In view of the equations  \eqref{e28}--\eqref{e30} and the quantized integral formulas stated in (\ref{e36}), we see that over the doubly periodic domain $\Omega$ the total `magnetic' flux is
  \be \label{flux}
-s\ito \mbox{Tr}(B)\, \ud x=2Na^2\left(\left[1+\sum\limits_{i=1}^{N-1}(R^{-1})_{1i}\right]\mu^2|\Omega|-\pi\sum\limits_{i=1}^{N-1}(R^{-1})_{1i}n_i\right)+
  2\pi\sum\limits_{i=1}^{N-1}(N-i)n_i,
\ee
which is not quantized and depends on $|\Omega|$, unless $a=0$. Here we have switched on the dependence
on the signature symbol $s=\pm1$ in the flux formula for generality. So the flux over the full plane $\mathbb{R}^2$ diverges
which leads to infinite energy as observed in \cite{kkkn}. Therefore, in order to avoid flux and energy divergence, it
is of value to develop an existence theory for doubly periodic solutions when $a>0$ and of
independent interest to spell out the existence 
theory separately when $a=0$. Indeed, when $a=0$, the matrix $R$ simplifies itself considerably so that the results
can be stated in concrete terms explicitly as follows.

   \begin{theorem}\label{tha2}
  For  $a=0,   \, \mu>0$, consider  the system of vortex equations \eqref{e28}--\eqref{e30} for  the field configuration
 $(f_1, \dots, f_{N-1})$ with the prescribed
  zero sets given  by  \eqref{e31} such that each $f_i$ has $n_i$ arbitrarily distributed zeros
$p_{i,1},\dots,p_{i,n_i}$, $i=1, \dots, N-1$.

  (i) For the problem  over the full plane $\mathbb{R}^2$,  there exists a  unique solution satisfying the boundary condition
  \be
  |f_i|^2\to (N-i),  \quad |x|\to \infty, \quad i=1, \dots, N-1.\label{e37}
  \ee
 Moreover, this  boundary condition is achieved exponentially fast at infinity
    \ber
    \left||f_i|^2-(N-i)\right| \le  C(\vep)\re^{-2\mu(1-\vep)\sqrt{\lambda_0}|x|}, \quad i=1, \dots, N-1, \label{e38}
     \eer
 where $\vep\in(0,1)$ is arbitrarily small, $C(\vep)$ is a positive constant
depending on $\vep$, and $\lambda_0$ is defined by \eqref{e32}.

(ii) For the problem  over a doubly periodic  domain $\Omega$,   there exists  a unique  solution if and only if
the conditions
  \ber
  \pi{\sum\limits_{j=i}^{N-1} \frac 1j\sum\limits_{l=1}^jn_l<\mu^2}|\Omega|(N-i), \quad i=1,\dots, N-1,\label{e39}
 \eer
 hold simultaneously.

 (iii) In both cases, there hold the quantized integrals
   \ber
    &&\int\left(|f_1|^2-|f_2|^2-1\right)\ud x=-\frac{\pi n_1}{\mu^2}, \label{e40} \\
    &&\int\left(-[i-1]|f_{i-1}|^2+[2i-1]|f_i|^2-i|f_{i+1}|^2-1\right)\ud x=-\frac{\pi n_i}{\mu^2}, \quad i=2, \dots, N-2,\label{e41}\\
    &&\int\left(-[N-2]|f_{N-2}|^2+[2N-3]|f_{N-1}|^2-1\right)\ud x=-\frac{\pi n_{N-1}}{\mu^2}, \label{e42}
  \eer
where the integration is evaluated either over the full plane $\mathbb{R}^2$ or the doubly periodic domain $\Omega$.
 \end{theorem}

To see the problem more transparently, we reformulate the system of
 equations \eqref{e28}--\eqref{e30} and Theorems \ref{tha1}--\ref{tha2}
in terms of a new family of  parameters and variables.
For  this purpose, we denote $N-1\equiv m$,  $\lambda\equiv4\mu^2$,  and set  \be 
u_i=\ln |f_i|^2, \quad i=1, \dots, m.\ee
 Then, in view of the  equations \eqref{e28}--\eqref{e30} and the zero sets  \eqref{e31}
we see that $u_1,\dots,u_m$ satisfy the equations
 \ber
  \Delta u_1&=&\lambda\left([2a^2+1]\re^{u_1}-[a^2+1]\re^{u_2}-1\right) +4\pi
  \sum\limits_{s=1}^{n_1}\delta_{p_{1,s}},\label{a2}\\
  \Delta u_i&=&\lambda\left(-[a^2+i-1]\re^{u_{i-1}}+[2a^2+2i-1]\re^{u_i}-[a^2+i]\re^{u_{i+1}}-1\right) +4\pi
  \sum\limits_{s=1}^{n_i}\delta_{p_{i,s}}, \label{a3}\\ && i= 2,\dots m-1,\nm\\
  \Delta u_m&=&\lambda\left(-[a^2+m-1]\re^{u_{m-1}}+[2a^2+2m-1]\re^{u_m}-1\right) +4\pi
  \sum\limits_{s=1}^{n_m}\delta_{p_{m,s}}.\label{a4}
 \eer

 Let
 \ber\left.\begin{array}{rrl}
 \mathbf{u}&=&(u_1, \dots, u_m)^\tau, \\
 \mathbf{1}&=&(1, \dots, 1)^\tau,\\
  \mathbf{U}&=&(\re^{u_1}, \dots, \re^{u_m})^\tau,\\
\mathbf{s}&=&\left(\sum\limits_{s=1}^{n_1}\delta_{p_{1,s}},\dots,\sum\limits_{s=1}^{n_m}\delta_{p_{m,s}}\right)^\tau.
\end{array}\right\}
\eer
Then the equations \eqref{a2}--\eqref{a4}  can be recast into its vector form
 \be
 \Delta\mathbf{u}=\lambda \left(R\mathbf{U}-\mathbf{1}\right)+4\pi
 \mathbf{s}.\label{a5}
 \ee

 Theorems \ref{tha1}--\ref{tha2} will be established through  proving the following results for  
the equations \eqref{a2}--\eqref{a4} or \eqref{a5}.

\begin{theorem}\label{th1}
  For any $a\ge0, \lambda>0$, consider  the system of equations \eqref{a2}--\eqref{a4}.

 (i) There exists a unique solution over $\mathbb{R}^2$  satisfying the boundary conditions
    \be
     u_i\to\ln\left(\sum\limits_{j=1}^m(R^{-1})_{ij}\right)\equiv \ln r_i,  \quad |x|\to \infty,\quad i=1,\dots,m. \label{a6}
    \ee
    Moreover, this solution satisfies the following exponential decay estimate at infinity:
    \ber
     \sum\limits_{i=1}^m(u_i(x)-\ln r_i)^2\le
     C(\vep)\re^{-(1-\vep)\sqrt{\lambda\lambda_0}|x|},\label{a7}
      \eer
 where $\vep\in(0,1)$ is arbitrarily small, $C(\vep)$ is a positive constant
depending on $\vep$, and $\lambda_0$ is defined by \eqref{e32}.

 (ii) For the problem over a doubly periodic domain $\Omega$, a solution exists 
 if and only if the following $m$ conditions
 \ber
  4\pi\sum\limits_{j=1}^m(R^{-1})_{ij}n_j<\lambda|\Omega|\sum\limits_{j=1}^m(R^{-1})_{ij}, \quad i=1, \dots, m, \label{a10}
 \eer
hold simultaneously. Besides, if a solution exists, it must be unique.

 (iii) In both full plane and periodic domain cases, there hold the  quantized integrals
  \be
   \int\left(\sum\limits_{j=1}^mR_{ij}\re^{u_j}-1\right)\ud x=-\frac{4\pi n_i}{\lambda}, \quad i=1, \dots, m, \label{a9}
  \ee
evaluated over the corresponding domain of consideration.
\end{theorem}

When $a=0$, we have the following  explicit results.

\begin{theorem}\label{th2}
   For $a=0$, $\lambda>0$, consider  the system of equations \eqref{a2}--\eqref{a4}.

 (i) There exists a unique solution over $\mathbb{R}^2$
 satisfying the boundary conditions
   \be
   u_i\to \ln(m-i+1),  \quad |x|\to \infty,\quad  i=1, \dots, m.\label{d4}
   \ee
     Moreover, this solution  satisfies the following exponential decay  estimate at infinity:
    \ber
     \sum\limits_{i=1}^m\left(u_i(x)-\ln(m-i+1)\right)^2\le  C(\vep)\re^{-(1-\vep)\sqrt{\lambda\lambda_0}|x|},\label{d5}
      \eer
 where $\vep\in(0,1)$ is small, $C(\vep)>0$ depends on $\vep$, and $\lambda_0$ is defined by \eqref{e32}.

  (ii) There exists a solution over a doubly periodic domain $\Omega$ if and only if the conditions
 \ber
   4\pi{\sum\limits_{j=i}^m \frac 1j\sum\limits_{l=1}^jn_l<\lambda}|\Omega|(m-i+1), \quad i=1,\dots, m,\label{d7}
 \eer
are fulfilled simultaneously. Besides, if a solution exists, it must be unique.

 (iii)  In both cases,  the quantized integrals
  \ber
    &&\int\left(\re^{u_1}-\re^{u_2}-1\right)\ud x=-\frac{4\pi n_1}{\lambda}, \\
    &&\int\left(-[i-1]\re^{u_{i-1}}+[2i-1]\re^{u_i}-i\re^{u_{i+1}}-1\right)\ud x=-\frac{4\pi n_i}{\lambda}, \quad i=1, \dots, m-1,\\
    &&\int\left(-[m-1]\re^{u_{m-1}}+[2m-1]\re^{u_m}-1\right)\ud x=-\frac{4\pi n_m}{\lambda},
  \eer
are valid over the corresponding domain of the problem.
\end{theorem}

In the subsequent sections, we  prove Theorems \ref{th1}--\ref{th2}.

\section{Variational principle and solution to planar case}
\setcounter{equation}{0}

In this section we establish the existence and uniqueness results for 
a solution of \eqref{a2}--\eqref{a4} over the full plane
and derive the stated decay estimates for the solution. Unlike the problems studied in \cite{liey1,liny1,liny2}
which can be readily formulated variationally, the problem here
needs more elaboration in order that its  hidden variational structure be unveiled. For this purpose, we
shall rely on the well-known Cholesky decomposition theorem for positive-definite matrices. The  variational structure
to be recognized
will allow us to prove the existence of a solution over the full plane as well as over a doubly periodic domain,
although the present section is devoted to the planar case. Below we split our study into a few subsections.

 \subsection{Cholesky decomposition for the matrix $R$}

First, we observe that the symmetric matrix $R$ is positive definite.  In fact, for any $a\ge0$, it is easy to check that each leading principal minor
 of $R$ is positive. That is,
    \be
   R_l\equiv\det ([R_{ij}]_{l\times l})>0, \quad i,j=1,\dots,l,\quad l=1, \dots, m. \label{b1}
    \ee

 By the Cholesky decomposition theorem \cite{goor} the matrix  $R$ can be expressed as the product of a lower triangular  matrix  $L$ and its
transpose, $R=LL^\tau, \,L=(L_{ij})_{m\times m}$.  Indeed, using the iteration scheme presented in \cite{goor}, that is,
\ber
\left.\begin{array}{rrl}\medskip
  L_{11}&=&\sqrt{R_{11}},\\
\quad L_{i1}&=&\frac{R_{i1}}{L_{11}}, \quad i=2, \dots, m,\\
  L_{ii}&=&\sqrt{R_{ii}-\sum\limits_{j=1}^{i-1}L^2_{ij}},\quad  i=2,\dots,  m,\\
  L_{ij}&=&\frac{R_{ij}-\sum\limits_{l=1}^{j-1}L_{ik}L_{jl}}{l_{jj}}, \quad i=j+1, \dots, m,\, j=2,\dots, m,
\end{array} \right\}
  \eer
we have
 \ber\left.\begin{array}{rrl}\medskip
  L_{11}&=&\sqrt{2a^2+1}=\sqrt{R_1},\quad  L_{21}=-\frac{(a^2+1)}{\sqrt{R_1}},\\
  L_{ii-1}&=&-(a^2+i-1)\sqrt{\frac{R_{i-2}}{R_{i-1}}},\quad
  L_{ii}=\sqrt{\frac{R_i}{R_{i-1}}},\quad i=2,\dots,m,\\
  L_{ij}&=&0, \quad1\le j<i-1,  \quad i=2, \dots, m.
 \end{array} \right\} \label{b2}
  \eer
Here and in the sequel,   we  follow the   convention $R_0=1$.  We have
  \begin{equation}
  L=\begin{pmatrix}
  \sqrt{R_1}& 0& 0&\dots&0\\
 -\frac{a^2+1}{\sqrt{R_1}} &\sqrt{\frac{R_2}{R_1}}& 0 &\dots&0\\
 0&-(a^2+2)\sqrt{\frac{R_1}{R_2}}&\sqrt{\frac{R_3}{R_2}}&\dots& 0\\
 \vdots& \vdots&\vdots&\ddots &\vdots\\
\vdots&\vdots&\vdots&\vdots&\vdots\\
 0&0& \cdots& -(a^2+m-1)\sqrt{\frac{R_{m-2}}{R_{m-1}}}&\sqrt{\frac{R_m}{R_{m-1}}}
 \end{pmatrix}.
 \end{equation}

Furthermore, a simple calculation enables us to find the lower triangular matrix $L^{-1}$ with
   \ber\left.\begin{array}{rrl}\medskip
  (L^{-1})_{11}&=&\frac{1}{\sqrt{R_1}},\\
  (L^{-1})_{ij}&=& \frac{(a^2+j)\cdots(a^2+i-1)R_{j-1}}{\sqrt{R_{i-1}R_i}},\quad  1\le j\le i-1, \quad i=2,\dots, m,\\
  (L^{-1})_{ii}&=&\sqrt{\frac{R_{i-1}}{R_i}},  \quad i=2,\dots, m.
  \end{array} \right\} \label{b3}
  \eer

By the expression of $L^{-1}$ we can  compute  the inverse of $R$ by the formula
  \be
   R^{-1}=(LL^\tau)^{-1}=(L^{-1})^\tau L^{-1},\label{b4}
  \ee
from which we can see directly that, for any $a\ge0$,  all entries  of $R^{-1}$ are  positive: $(R^{-1})_{ij}>0, \,i,j=1,\dots, m$, as claimed earlier.

 \subsection{Variational formulation}
Following \cite{jata}, we introduce the background functions
 \be
  u_0^i=-\sum\limits_{s=1}^{n_i}\ln(1+\nu|x-p_{i,s}|^{-2}), \quad i=1, \dots, m, \label{b5}
 \ee
with $\nu>0$  being a parameter which should not be confused with the Lorentzian index used earlier. We see
that $u_0^i$ satisfy
 \be
  \Delta u_0^i=4\pi\sum\limits_{s=1}^{n_i}\delta_{p_{i,s}}-g_i,
\quad g_i=\sum\limits_{s=1}^{n_i}\frac{4\nu}{(\nu+|x-p_{i,s}|^2)^2}  \quad i=1, \dots, m.\label{b6}
 \ee
It is easy to see that
  \be
 g_i\in L(\mathbb{R}^2)\cap L^2(\mathbb{R}^2),\quad \itr g_i\ud x=4\pi n_i, \quad i=1, \dots, m. \label{b7}
  \ee
Since for any $a\ge 0$ each entry of $R^{-1}$ is positive, we have
 \be
  r_i\equiv\sum\limits_{j=1}^m(R^{-1})_{ij}>0, \quad i=1, \dots, m. \label{b8}
 \ee
Set
$\mathbf{r}=(r_1, \dots, r_m)^\tau$
and 
$
  u_i=u_0^i+\ln r_i+v_i,  i=1, \dots m$, $\mathbf{v}=(v_1, \dots, v_m)^\tau,$ $\mathbf{g}=(g_1, \dots,g_m)^\tau$,
$ \mathbf{U}=(\re^{u_0^1+v_1}, \dots, \re^{u_0^m+v_m})^\tau$.
 Then the equations \eqref{a2}--\eqref{a4} over $\mathbb{R}^2$ become
 \ber
  \Delta v_1&=&\lambda \left([2a^2+1]r_1\re^{u_0^1+v_1}-[a^2+1]r_2\re^{u^2_0+v_2}-1\right)+g_1,\label{b9}\\
  \Delta v_i&=&\lambda \left(-[a^2+i-1]r_{i-1}\re^{u_0^{i-1}+v_{i-1}}+[2a^2+2i-1]r_i\re^{u_0^i+v_i}\right.\nm\\
&&\left.-[a^2+i]r_{i+1}\re^{u_0^{i+1}+v_{i+1}}-1\right)
  +g_i,\quad  i= 2,\dots m-1,\label{b10}\\
  \Delta v_m&=&\lambda \left(-[a^2+m-1]r_{m-1}\re^{u_0^{m-1}+v_{m-1}}+[2a^2+2m-1]r_m\re^{u_0^m+v_m}-1\right) +g_m. \label{b11}
 \eer
Or in its equivalent  vector form, we have
 \be
 \Delta\mathbf{v}=\lambda R\,\mathrm{diag}\{r_1, \dots, r_m\}\left(\mathbf{U}-\mathbf{1}\right)+\mathbf{g},\label{b12}
 \ee
since $R\,\mathrm{diag}\{r_1, \dots, r_m\}{\bf 1}={\bf 1}$ by the definition of $r_1,\dots, r_m$.
Now we use the notation
  \be
  \quad \mathbf{w}=(w_1, \dots,  w_m)^\tau,\quad
   \mathbf{h}=\frac1\lambda L^{-1}\mathbf{g}, \quad
   \mathbf{h}=(h_1,\dots,h_m)^\tau \label{b13}
  \ee
  and the transformation
\begin{equation}\label{b14}
\mathbf{w}=L^{-1}\mathbf{v}=\begin{pmatrix}
  (L^{-1})_{11}v_1\\
(L^{-1})_{21}v_1+(L^{-1})_{22}v_2\\
 \vdots\\
 (L^{-1})_{m1}v_1+\dots+(L^{-1})_{mm}v_m
\end{pmatrix},
\ee
or
\be
  \mathbf{v}=L\mathbf{w}=
  \begin{pmatrix}L_{11}w_1\\
   L_{21}w_1+L_{22}w_2\\
   \vdots\\
   L_{mm-1}w_{m-1}+L_{mm}w_m
  \end{pmatrix}.\label{b15}
   \ee
Then the equations \eqref{b9}--\eqref{b11} take the form:
 \ber
 \Delta w_1&=&\lambda\left(L_{11}r_1\left[\re^{u_0^1+L_{11}w_1}-1\right]+L_{21}r_2\left[\re^{u_0^2+L_{21}w_1+L_{22}w_2}-1\right]+h_1\right)\label{b16}\\
 \Delta w_i&=&\lambda\left(L_{ii}r_i\left[\re^{u_0^i+L_{ii-1}w_{i-1}+L_{ii}w_i}-1\right]\right.\nm\\
&&\left.+L_{i+1i}r_{i+1}\left[\re^{u_0^{i+1}+L_{i+1i}w_i+L_{i+1i+1}w_{i+1}}-1\right]
+h_i\right), \quad i=2, \dots, m-1, \label{b17} \\
 \Delta w_m&=&\lambda\left( L_{mm}r_m\left[\re^{u_0^m+L_{mm-1}w_{m-1}+L_{mm}w_m}-1\right]+h_m\right).\label{b18}
 \eer
Or in its  vector form, we have
 \be
 \Delta\mathbf{w}=\lambda \left(L^\tau\mathrm{diag}\{r_1,  \dots, r_m\}\left[\mathbf{U}-\mathbf{1}\right]+\mathbf{h}\right).\label{b19}
 \ee

It can now be checked to  see that the equations \eqref{b16}--\eqref{b18} or \eqref{b19}  are the Euler--Lagrange equations of the functional
  \ber
  I(\mathbf{w})&=&\frac{1}{2\lambda}\sum\limits_{i=1}^m\itr|\nabla w_i|^2\ud x+ \sum\limits_{i=1}^m\itr h_iw_i\ud x+\itr r_1\left(\re^{u_0^1+L_{11}w_1}-\re^{u_0^1}-L_{11}w_1\right)\ud x\nm\\
   &&+\sum\limits_{i=2}^m\itr r_i\left(\re^{u_0^i+L_{ii-1}w_{i-1}+L_{ii}w_i}-\re^{u_0^i}-\left[L_{ii-1}w_{i-1}+L_{ii}w_i\right]\right)\ud x.\label{b20}
  \eer
This is the variational principle we have aimed to unveil.

To facilitate the computation and analysis, it will be technically more convenient to rewrite the functional (\ref{b20}) as
   \ber
   I(\mathbf{w}) &=&\frac{1}{2\lambda}\sum\limits_{i=1}^m\itr|\nabla w_i|^2\ud x+\itr r_1\re^{u_0^1}\left(\re^{L_{11}w_1}-1-L_{11}w_1\right)\ud x\nm\\
   &&+\sum\limits_{i=2}^m\itr r_i\re^{u_0^i}\left(\re^{L_{ii-1}w_{i-1}+L_{ii}w_i}-1-\left[L_{ii-1}w_{i-1}+L_{ii}w_i\right]\right)\ud x\nm\\
   && +\sum\limits_{i=1}^{m-1}\itr\left(L_{ii}r_i\left[\re^{u_0^i}-1\right]+L_{i+1i}r_{i+1}\left[\re^{u_0^{i+1}}-1\right]+h_i\right)w_i\ud x\nm\\
  &&+\itr\left(L_{mm}r_m[\re^{u_0^m}-1]+h_m\right)w_m\ud x,\label{b21}
  \eer
which allows us to approach the problem in a similar manner as in \cite{jata,T1} for the scalar situation, as we
will do in the following.

To proceed, we can compute to get
  \ber
  &&(DI(\mathbf{w}))(\mathbf{w})\nm\\
  &&=\frac{1}{\lambda}\sum\limits_{i=1}^m\itr|\nabla w_i|^2\ud x+\itr\left(L_{11}r_1\re^{u_0^1}\left[\re^{L_{11}w_1}-1\right]+L_{21}r_2\re^{u_0^2}\left[\re^{L_{21}w_1+L_{22}w_2}-1\right]\right)w_1 \ud x\nm\\
  &&+\sum\limits_{i=2}^{m-1}\itr\left(L_{ii}r_i\re^{u_0^i}\left[\re^{L_{ii-1}w_{i-1}+L_{ii}w_i}-1\right]+L_{i+1i}r_{i+1}\re^{u_0^{i+1}}\left[\re^{L_{i+1i}w_i+L_{i+1i+1}w_{i+1}}-1\right]\right)w_i \ud x\nm\\
  &&+\itr\left(L_{mm}r_m\re^{u_0^m}\left[\re^{L_{mm-1}w_{m-1}+L_{mm}w_m}-1\right]\right)w_m \ud x\nm\\
  &&+\sum\limits_{i=1}^{m-1}\itr \left(L_{ii}r_i\left[\re^{u_0^i}-1\right]+L_{i+1i}r_{i+1}\left[\re^{u_0^{i+1}}-1\right]+h_i\right)w_i\ud x \nm \\
  &&+\itr \left(L_{mm}r_m\left[\re^{u_0^m}-1\right]+h_m\right)w_m\ud x\nm\\
  &&=\frac{1}{\lambda}\sum\limits_{i=1}^m\itr|\nabla w_i|^2\ud x+\itr r_1\left(\re^{u_0^1+L_{11}w_1}-1+\tilde{h}_1\right)L_{11}w_1\ud x\nm\\
  &&+\sum\limits_{i=2}^m\itr r_i\left(\re^{u_0^i+L_{ii-1}w_{i-1}+L_{ii}w_i}-1+\tilde{h}_i\right)\left(L_{ii-1}w_{i-1}+L_{ii}w_i\right)\ud x,\label{b22}
  \eer
where $\tilde{h}_i$'s are some  linear combinations of $h_i$'s, also of  $g_i$'s. More precisely,
 \be 
\tilde{h}_i=\frac{1}{\lambda r_i}\sum\limits_{j=1}^m(R^{-1})_{ji}g_j, \quad i=1,\dots,m.
\ee

Using the invertibility of the transformations \eqref{b14} and \eqref{b15},  we see that there exist  some positive constants $c_1$ and $c_2$ such that
  \ber
   c_1\sum\limits_{i=1}^mv_i^2\le\sum\limits_{i=1}^mw_i^2\le c_2\sum\limits_{i=1}^mv_i^2, \quad
   c_1\sum\limits_{i=1}^m|\nabla v_i|^2\le\sum\limits_{i=1}^m|\nabla w_i|^2\le c_2\sum\limits_{i=1}^m|\nabla v_i|^2. \label{b23}
  \eer
Therefore, from \eqref{b15}, \eqref{b22} and \eqref{b23}, we can obtain
 \ber
 (DI(\mathbf{w}))(\mathbf{w}) \ge C_0\sum\limits_{i=1}^m\itr|\nabla v_i|^2\ud x+\sum\limits_{i=1}^m\itr r_i\left(\re^{u_0^i+v_i}-1+\tilde{h}_i\right)v_i\ud x,  \label{b24}
  \eer
 where $C_0$ is a  positive constant.

To deal with the second term on the right hand side of \eqref{b24}, we follow the approach of \cite{jata}. We just need to estimate a typical  term of the following  form
  \be
  M(v)=\itr\left(\re^{u_0+v}-1+\tilde{h}\right)v\ud x.
  \ee
It is easy to see that $M(v)$ can be decomposed as
  $M(v)=M(v_+)+M(-v_-)$
where $v_+=\max\{v, \,0\}, v_-=\max\{-v, \,0\}$.

From the elementary inequality $\re^t-1\ge t, \,t\in \mathbb{R}$ and the fact $u_0,\, \tilde{h}\in L^2(\mathbb{R}^2)$,  we have
 \ber
 M(v_+)\ge\itr(u_0+v_++\tilde{h})v_+\ud x\ge \frac12\itr v_+^2\ud x-C_1, \label{b25}
 \eer
 where and  in the sequel  we use  $C$ to denote a generic positive constant.

To estimate $M(-v_-)$, we note the inequality $1-\re^{-t}\ge\frac{t}{1+t},\, t\ge0$,  and obtain
 \ber
   M(-v_-)&=&\itr\left(1-\re^{u_0-v_-}-\tilde{h}\right)v_-\ud x\nm\\
   &=&\itr\left(1-\tilde{h}-\re^{u_0}+\re^{u_0}\left[1-\re^{-v_-}\right]\right)v_-\ud x\nm\\
   &\ge&\itr\left(1-\tilde{h}-\re^{u_0}+\re^{u_0}\frac{v_-}{1+v_-}\right)v_-\ud x\nm\\
   &=&\itr\left([1+v_-]\left[1-\tilde{h}-\re^{u_0}\right]+\re^{u_0}v_-\right)\frac{v_-}{1+v_-}\ud x\nm\\
   &=&\itr\left(1-\tilde{h}\right)\frac{v_-^2}{1+v_-}\ud x+\itr \left(1-\re^{u_0}-\tilde{h}\right)\frac{v_-}{1+v_-}\ud x \nm\\
   &\ge&\frac12\itr\frac{v_-^2}{1+v_-}\ud x+\itr \left(1-\re^{u_0}-\tilde{h}\right)\frac{v_-}{1+v_-}\ud x \nm\\
   &\ge&\frac14 \itr\frac{v_-^2}{(1+v_-)^2}\ud x-C_2,\label{b26}
 \eer
 where we have used  $\tilde{h}\le \frac12$, assured by taking $\nu$ sufficiently large, and the fact
that both $\re^{u_0}-1$ and $\tilde{h}$ belong to $L^2(\mathbb{R}^2)$.

Then we see from  \eqref{b25} and \eqref{b26} that
 \be
  M(v)\ge \frac14\itr\frac{v^2}{(1+|v|)^2}\ud x-C. \label{b27}
 \ee

 To proceed further, we  need  the following standard interpolation inequality over $\wot$:
 \be
 \itr v^4\ud x\le 2\itr v^2\ud x\itr |\nabla v|^2\ud x, \quad \forall\, v\in \wot. \label{b28}
 \ee
Using \eqref{b28}, we have
  \ber
   &&\left(\itr|v|^2\ud x\right)^2\nm\\
   &&=\left(\itr\frac{|v|}{1+|v|}(1+|v|)|v|\ud  x\right)^2\nm\\
    &&\le\itr\frac{|v|^2}{(1+|v|)^2}\ud x\itr\big(|v|+|v|^2\big)^2\ud x\nm\\
    &&\le4\itr\frac{|v|^2}{(1+|v|)^2}\ud x\itr|v|^2\ud x\left(\itr|\nabla v|^2\,\ud x+1\right)\nm\\
    &&\le\frac12 \left(\itr|v|^2\ud x\right)^2+C\left(\left[\frac{|v|^2}{(1+|v|)^2}\ud x\right]^4 +\left[\itr|\nabla v|^2\ud x\right]^4+1\right),
  \eer
  which  implies
   \ber
   \|v\|_2&\le& C\left(\itr \frac{|v|^2}{(1+|v|)^2}\ud x+\itr|\nabla v|^2\ud x+1\right), \label{b29}
   \eer
where and in the sequel we use $\|\cdot\|_p$ to denote the norm of the space $L^p(\mathbb{R}^2)$.

From \eqref{b24} and \eqref{b27}, we obtain
  \be
   (DI(\mathbf{w}))(\mathbf{w})\ge C_2\sum\limits_{j=1}^m\itr\left(|\nabla v_j|^2+\frac{v^2_j}{(1+|v_j|)^2}\right)\ud x-C_3 .\label{b30}
  \ee
Then it follows from \eqref{b23}, \eqref{b29} and \eqref{b30} that
  \ber
  (DI(\mathbf{w}))(\mathbf{w})\ge  C_4\sum\limits_{j=1}^m\|w_j\|_{\wot}-C_5.\label{b31}
  \eer

By the coercive lower bound  in \eqref{b31}, we can show  that the functional $I$ defined by \eqref{b20} admits a critical point.  In fact, by \eqref{b31}, we can choose $\xi>0$ such that
   \be
    \inf\{DI(\mathbf{w})(\mathbf{w}) \,|\,\|\mathbf{w}\|_{\wot}=\xi \}\ge 1.
   \ee
Since  the functional $I$ is weakly  lower semi-continuous on $\wot$, the minimization problem
  \be
  \eta_0\equiv\inf\left\{I(\mathbf{w})|\, \|\mathbf{w}\|_{\wot}\le  \xi\right\} \label{b32}
  \ee
admits a solution, say, $\tilde{\mathbf{w}}$. Then, we can show that it must be an interior point. We argue by contradiction. Assume that $\|\tilde{\mathbf{w}}\|_{\wot}= \xi$. Then
 \be
  \lim\limits_{t\to0}\frac{I((1-t)\tilde{\mathbf{w}})-I(\tilde{\mathbf{w}})}{t}
  = \frac{\ud }{\ud t}I((1-t)\tilde{\mathbf{w}})\big|_{t=0}=-(DI(\tilde{\mathbf{w}}))(\tilde{\mathbf{w}})\le -1
 \ee
Therefore, if $t>0$ is sufficiently small, letting $\tilde{\mathbf{w}}^t=(1-t)\tilde{\mathbf{w}}$, we can obtain
 \be 
I(\tilde{\mathbf{w}}^t)<I(\tilde{\mathbf{w}})=\eta_0, \quad \|\tilde{\mathbf{w}}^t\|_{\wot}=(1-t)\xi<\xi,\ee
which contradicts the definition of $\eta_0$. Hence, $\tilde{\mathbf{w}}$ must be an interior critical  point for the
problem \eqref{b32}. As a result, it is a critical point of the functional $I$. Since the functional $I$  is strictly convex, this critical point must be unique.

\subsection{ Asymptotic behavior of the solution at infinity}
Here we study the asymptotic behavior of the solution obtained above. Noting that  $\mathbf{w}\in \wot$, by the well-known inequality
  \ber
   \|\re^v-1\|_2^2\le C_1\exp(C_2\|v\|_{\wot}^2),\quad \forall \,v\in \wot,
  \eer
where $C_1, C_2$ are some positive constants, we see that the right-hand sides of the equations \eqref{b16}--\eqref{b18} all belong to $L^2(\mathbb{R}^2)$. Using the
standard elliptic $L^2$-estimates, we have $w_j\in W^{2,2}(\mathbb{R}^2)$, which implies $w_j\to 0$ as
$|x|\to \infty$, $j=1,\dots, m$. By the transformation \eqref{b15}, we see that $v_j\to 0$ as $|x|\to \infty$, which implies the desired
boundary condition $u_j\to \ln r_j$ as $|x|\to \infty, \, j=1, \dots, m$.

 Next we show that $|\nabla w_j|\to0$ as $|x|\to\infty$, $j=1,\dots,m$.  A typical term of the right hand sides of \eqref{b16}-\eqref{b18} can be rewritten as
  \ber
   \re^{u_0^j+L_{jj-1}w_{j-1}+L_{jj}w_j}-1=\left(\re^{u_0^j}-1\right)\re^{L_{jj-1}w_{j-1}+L_{jj}w_j}+\left(\re^{L_{jj-1}w_{j-1}+L_{jj}w_j}-1\right),
  \eer
which belongs to $L^p(\mathbb{R}^2)$ for any $p>2$ due to the embedding $\wot\subset L^p(\mathbb{R}^2)$ and the definition of
$u_0^j$.  Therefore all the right-hand-side terms of \eqref{b16}-\eqref{b18} lie in $L^p(\mathbb{R}^2)$, for any $p>2$.
Then the elliptic $L^p$-estimates imply $w_j\in W^{2,p}(\mathbb{R}^2)$ for any $p>2$, $j=1,\dots, m$. Consequently,
we have $|\nabla w_j|\to 0$ as $|x|\to \infty$, $j=1,\dots m$. That is,  $|\nabla u_j|\to 0$ as $|x|\to \infty$, $j=1,\dots m$.

Now we   establish the exponential decay rate of the solutions at infinity. To this end, we consider the equations
\eqref{a2}-\eqref{a4} or \eqref{a5} over an exterior domain  \be D_\rho=\left\{x\in \mathbb{R}^2| \quad |x|>\rho\right\},\ee
where $\rho>0$ satisfies \be\rho>\max\big\{ |p_{i,s}|\, | \, i=1,\dots,m,\,s=1,\dots,n_i\big\}.\ee

For convenience, we consider the system of equations \eqref{a5} over $D_\rho$. Recall that
${\bf r}=R^{-1}{\bf 1}$. Hence we may rewrite (\ref{a5}) in $D_\rho$ as
\be \label{345}
\Delta {\bf u}=\lambda R({\bf U}-{\bf r})=\lambda R{\bf v}+\lambda R({\bf U}-{\bf r}-{\bf v}),
\ee
where vector $\bf v$ is to be determined. 

Since the matrix $R$ is positive definite,  there is  an
orthogonal matrix $O$ such that \be O^\tau RO={\rm diag}\{\lambda_1,\dots, \lambda_m\},\quad \min\{\lambda_1,\dots,
\lambda_m\}>0.\ee

Now apply $O^\tau$ in (\ref{345}) and set
 \be\tilde{\mathbf{u}}=O^\tau{\bf v},\quad {\bf v}=(u_1-\ln r_1, \dots, u_m-\ln r_m)^\tau.\ee 
Then we have
 \ber
 \Delta \tilde{\mathbf{u}}=\lambda{\rm diag}\{\lambda_1,\dots, \lambda_m\}\tilde{\mathbf{u}} + 
\lambda O^\tau R({\bf U}-{\bf r}-{\bf v}). \label{b34}
 \eer
Note that, since ${\bf U}\to {\bf r}$ as $|x|\to\infty$, we have ${\bf U}-{\bf r}=E(x){\bf v}$ where $E(x)$ is an
$m\times m$ diagonal
matrix so that $E(x)\to I_m$ (the $m\times m$ identity matrix) as $|x|\to\infty$. This observation leads us to rewrite
 \eqref{b34} as
\be \label{P}
\Delta \tilde{\mathbf{u}}=\lambda{\rm diag}\{\lambda_1,\dots, \lambda_m\}\tilde{\mathbf{u}} +\lambda P(x)\tilde{\bf u},
\ee
where $P(x)$ is an $m\times m$ matrix which vanishes at infinity. As a consequence of (\ref{P}), we obtain
  \ber
  \Delta|\tilde{\mathbf{u}}|^2\ge 2\tilde{\mathbf{u}}^\tau\Delta
  \tilde{\mathbf{u}}\ge \lambda
  \lambda_0|\tilde{\mathbf{u}}|^2- b(x)|\tilde{\mathbf{u}}|^2
  \eer
with $b(x)\to 0$ as $|x|\to \infty$.

Then, for any $\vep\in (0,1)$, we can find a suitably large $\rho_\vep\geq\rho$ such that
 \be \label{u}
\Delta|\tilde{\mathbf{u}}|^2\ge\left(1- \frac{\vep}{2}\right)\lambda\lambda_0|\tilde{\mathbf{u}}|^2, \quad x\in D_{\rho_\vep}.\ee

We will use a comparison function, say $\eta$, of the form
 \be\label{52}
\eta = C\re^{-\sigma|x|},\quad |x|>0,\quad C,\sigma\in\mathbb{R},\quad C,\sigma>0.
\ee
Then $\Delta\eta=\sigma^2\eta-\frac\sigma{|x|}\eta$. Thus, in view of (\ref{u}), we have
\be 
\Delta\left(|\tilde{\bf u}|^2-\eta\right)\geq\left(1-\frac\vep2\right)\lambda\lambda_0|\tilde{\bf u}|^2-\sigma^2\eta,\quad |x|\geq \rho_\vep.
\ee
We take the obvious choice $\sigma^2=\left(1-\frac\vep2\right)\lambda\lambda_0$ which gives us
$\Delta(|\tilde{\bf u}|^2-\eta)\geq\sigma^2(|\tilde{\bf u}|^2-\eta)$, $|x|\geq\rho_\vep$.
Choose $C$ in (\ref{52}) large so that $|\tilde{\bf u}|^2-\eta\leq 0$ for $|x|=\rho_\vep$.
Hence, using the fact that $|\tilde{\mathbf{u}}|\to 0$ as $|x|\to \infty$  and
the maximum principle, we see that $|\tilde{\bf u}|^2\leq\eta$ for $|x|\geq\rho_\vep$. So the estimate

   \be |\tilde{\mathbf{u}}|^2\le C(\vep)\re^{-(1-\vep)\sqrt{\lambda\lambda_0}|x|},\quad |x|\geq\rho_\vep,\ee
 follows since $(1-\vep)^2<\left(1-\frac\vep2\right)$
for any $\vep\in(0,1)$. 
Therefore the desired exponential decay rate \eqref{a7} is established.

To get the quantized integrals, we need to  establish the exponential decay rate for the derivatives.  

Let $\partial$ denote any of the two derivatives $\partial_1$ and  $\partial_2$.  Define
 \be
{\mathbf{v}}=(\partial u_1, \dots, \partial u_m)^\tau, \quad P={\rm diag}\{r_1, \dots, r_m\},\quad Q={\rm diag}\{\re^{u_1}-r_1, \dots, \re^{u_m}-r_m\}.
 \ee
Then differentiating  \eqref{a5} in $D_\rho$, we have
  \be
   \Delta {\mathbf{v}}=\lambda RP{\mathbf{v}}+\lambda  RQ{\mathbf{v}}.\label{b35}
  \ee
Let $O$ be as before and set \be P{\mathbf{v}}=O \tilde{\mathbf{u}}, \quad f={\mathbf{v}}^\tau P{\mathbf{v}}.\ee
  Then by \eqref{b35} and the fact that $r_1,\dots,r_m>0$, we obtain
   \ber
   \Delta f&\ge& 2{\mathbf{v}}^\tau P\Delta{\mathbf{v}}= 2\lambda{\mathbf{v}}^\tau PRP{\mathbf{v}}+2\lambda{\mathbf{v}}^\tau PRQ{\mathbf{v}}\nm\\
   &\ge&\lambda\lambda_0|\tilde{\mathbf{u}}|^2+2\lambda{\mathbf{v}}^\tau PRQ{\mathbf{v}}= \lambda\lambda_0{\mathbf{v}}^\tau P^2{\mathbf{v}}+2\lambda{\mathbf{v}}^\tau PRQ{\mathbf{v}}\nm\\
   &\ge& \lambda \lambda_0r_0f-b(x)f,
   \eer
where $r_0=\min\{r_1, \dots, r_m\}$ and $b(x)\to 0$ as $|x|\to\infty$. Then, as discussed previously, 
we can conclude that for any $\vep\in(0,1)$, there is a positive constant $C(\vep)>0$, such that
  \be 
f\le C(\vep)\re^{-(1-\vep)\sqrt{\lambda\lambda_0r_0}|x|},\ee
when $|x|$ is sufficiently large. Thus  we get the following exponential decay rate near infinity:
 \be
  \sum\limits_{i=1}^m|\nabla u_i(x)|^2\le C(\vep)\re^{-(1-\vep)\sqrt{\lambda\lambda_0r_0}|x|}.\label{a8}
 \ee

\medskip

We can now calculate  the quantized integrals \eqref{a9} stated in Theorem \ref{th1} for the planar case.

Using \eqref{b5}, \eqref{b6}, and  the exponential decay property  of $|\nabla u_i|$'s in \eqref{a8}, we conclude that $|\nabla  v_i|$'s vanish
at infinity at least at the rate $|x|^{-3}$. Thus, using the divergence theorem, we have
\be \label{div}
\int_{\mathbb{R}^2}\Delta v_i\,\ud x=0,\quad i=1,\dots,m.
\ee
Consequently, integrating the equations \eqref{b9}--\eqref{b11} over $\mathbb{R}^2$,   
and applying \eqref{b7} and (\ref{div}), we  obtain the desired results stated in \eqref{a9}.

We next turn our attention to the compact case.

\section{Doubly periodic case}
\setcounter{equation}{0}
 In this section we  consider the equations \eqref{a2}--\eqref{a4} over a doubly periodic domain $\Omega$.

Let $u_0^i$ be a solution of the problem (see\cite{aubi})
 \ber
  \Delta u_0^i=-\frac{4\pi n_i}{|\Omega|}+4\pi\sum\limits_{s=1}^{n_i}\delta_{p_{i,s}}, \quad  x\in \Omega,\quad i=1, \dots, m.
 \eer

Set  $u_i=u_0^i+v_i, i=1, \dots, m$. Then equations  \eqref{a2}--\eqref{a4} become
 \ber
 \Delta v_1&=&\lambda\left(\left[2a^2+1\right]\re^{u_0^1+v_1}-\left[a^2+1\right]\re^{u_0^2+v_2}-1\right)+\frac{4\pi n_1}{|\Omega|},\label{c1} \\
 \Delta v_i&=&\lambda\left(-\left[a^2+i-1\right]\re^{u_0^{i-1}+v_{i-1}}+\left[2a^2+2i-1\right]\re^{u_0^i+v_i}-\left[a^2+i\right]\re^{u_0^{i+1}+v_{i+1}}-1\right)\nm \\
 &&+\frac{4\pi n_i}{|\Omega|}, \quad i=2, \dots,m-1, \label{c2}\\
 \Delta v_m&=&\lambda\left(-\left[a^2+m-1\right]\re^{u_0^{m-1}+v_{m-1}}+\left[2a^2+2m-1\right]\re^{u_0^m+v_m}-1\right)+\frac{4\pi n_m}{|\Omega|},\label{c3}
\eer
 or equivalently in its vector form
  \ber
   \Delta \mathbf{v}=\lambda(R\mathbf{U}-\mathbf{1})+\frac{4\pi}{|\Omega|}\mathbf{n},  \label{c4}
  \eer
with
$\mathbf{U}=(\re^{u_0^1+v_1}, \dots, \re^{u_0^m+v_m})^\tau, \mathbf{v}=(v_1, \dots v_m)^\tau,   \mathbf{1}=(1,\dots, 1)^\tau,  \mathbf{n}=(n_1, \dots, n_m)^\tau.$

Integrating the equations \eqref{c1}--\eqref{c3} or \eqref{c4} over $\Omega$, we can obtain the natural constraints
   \ber
    R\ito \mathbf{U}\ud x =|\Omega|\mathbf{1}-\frac{4\pi}{\lambda}\mathbf{n}.  \label{c5}
   \eer
 We may rewrite (\ref{c5}) more conveniently as
   \be
    \ito \mathbf{U}\ud
    x=|\Omega|R^{-1}\mathbf{1}-\frac{4\pi}{\lambda}R^{-1}\mathbf{n}\equiv\mathbf{K}, \label{c6}
   \ee
or in its component form
   \ber
    \ito \re^{u_0^i+v_i}\,\ud x=|\Omega|\sum\limits_{j=1}^m(R^{-1})_{ij}-\frac{4\pi}{\lambda}\sum\limits_{j=1}^m(R^{-1})_{ij}n_j= K_i, \quad i=1,\dots, m,  \label{c7}
    \eer
where and in what follows we use the notation
 $\mathbf{K}=(K_1, \dots, K_m)^\tau.$
Therefore, we see that if  a solution exists, then $K_i>0, i=1,\dots, m$. As a result, we get the necessity of the condition \eqref{a6}.

In what follows we show that the condition \eqref{a6} is also sufficient for the existence of a
solution to  \eqref{c1}--\eqref{c3} by variational methods.

We will work on the Sobolev spaces $W^{1,2}(\Omega)$, which is composed of scalar- or vector-valued $\Omega$-periodic $L^2$-functions  whose derivatives also  belong to  $L^2(\Omega)$. For the
scalar case we have the decomposition
  \be 
W^{1,2}(\Omega)=\mathbb{R}\oplus\dot{W}^{1, 2}(\Omega),\ee
 where
\be\dot{W}^{1,2}(\Omega)=\left\{w\in W^{1,2}(\Omega)\Bigg| \ito w\ud x=0\right\},\ee
  is a closed subspace of $W^{1,2}(\Omega)$. Then for any $u\in W^{1, 2}(\Omega)$, we have
 \be
 u=\underline{u}+\dot{u}, \quad \underline{u}\in \mathbb{R}, \quad \dot{u}\in \dot{W}^{1,2}(\Omega).\label{c7'}
 \ee

 We will use the well-known   Trudinger--Moser inequality  \cite{aubi,font}
 \be
 \ito \re^{u}\ud x\le C\exp{\left(\frac{1}{16\pi}\ito|\nabla u|^2\ud x\right)}, \quad \forall\, u\in\dot{W}^{1,2}(\Omega), \label{c8}
 \ee
which is important for our estimate, although the analysis does not depend on the optimality of the embedding constant.

 As in the planar case, to formulate the problem in a variational structure, we use the transformation \eqref{b14} or \eqref{b15}.
Then the equations \eqref{c1}--\eqref{c3}  become
 \ber
 \Delta w_1&=&\lambda\left(L_{11}\re^{u_0^1+L_{11}w_1} +L_{21}\re^{u_0^2+L_{21}w_1+L_{22}w_2}-b_1\right),\label{c9}\\
 \Delta w_i&=&\lambda\left(L_{ii}\re^{u_0^i+L_{ii-1}w_{i-1}+L_{ii}w_i}+L_{i+1i}\re^{u_0^{i+1}+L_{i+1i}w_i+L_{i+1i+1}w_{i+1}}-b_i\right), \label{c10} \\
 &&i=2, \dots, m-1\nm \\
 \Delta w_m&=&\lambda \left(L_{mm} \re^{u_0^m+L_{mm-1}w_{m-1}+L_{mm}w_m}-b_m\right),\label{c11}
 \eer
whose  vector form is
 \be
 \Delta\mathbf{w}=\lambda \left(L^\tau\mathbf{U}-\mathbf{b}\right),\label{c12}
 \ee
where
 \ber
 \mathbf{b}=(b_1, \dots, b_m)^\tau\equiv L^{-1}\mathbf{1}-\frac{4\pi}{\lambda |\Omega|}L^{-1}\mathbf{n}, \label{c13}
 \eer

As before, we can check to see that the equations \eqref{c9}--\eqref{c11} are the Euler--Lagrange equations of the functional
  \be
  I(\mathbf{w})=\frac{1}{2\lambda}\sum\limits_{i=1}^m\ito|\nabla w_i|^2\ud x +\ito\left(\re^{u_0^1+L_{11}w_1}+ \sum_{i=2}^m \re^{u_0^i+L_{ii-1}w_{i-1}+L_{ii}w_i}\right)\ud x 
  -\sum_{i=1}^m\ito b_iw_i\ud x.\label{c14}
  \ee

 We shall now engage ourselves in a direct minimization procedure, initiated in  \cite{liey1},  to find a critical point of the functional (\ref{c14}).

When $\mathbf{w}\in W^{1,2}(\Omega)$, by the Trudinger--Moser inequality \eqref{c8}, we  see that the functional defined by \eqref{c14} is a $C^1$-functional
and lower semi-continuous with respect to the weak topology of $W^{1, 2}(\Omega)$.

Using the decomposition formula \eqref{c7'}, we  obtain
 \ber
  I(\mathbf{w})-\frac{1}{2\lambda}\sum\limits_{i=1}^m\|\nabla\dot{w}_i\|_2^2
   &=&\ito \left(\re^{u_0^1+L_{11}w_1}+\sum\limits_{i=2}^m\re^{u_0^i+L_{ii-1}w_{i-1}+L_{ii}w_i}\right)\ud  x- |\Omega|\mathbf{b}^\tau\underline{\mathbf{w}}\nm\\
   &=&\sum\limits_{i=1}^m\ito\re^{u_0^i+v_i}\ud x -\mathbf{K}^\tau\underline{\mathbf{v}}\nm\\
   &=& \sum\limits_{i=1}^m\ito\re^{u_0^i+\dot{v}_i+\underline{v}_i}\ud x -\sum\limits_{i=1}^mK_i\underline{v}_i,\label{c16}
 \eer
where  we have  used
$ |\Omega|\mathbf{b}^\tau\underline{\mathbf{w}}=\mathbf{K}^\tau\underline{\mathbf{v}}$
in view of \eqref{c6} and \eqref{c13}.

By  Jensen's inequality, we see that
 \ber
 \ito \re^{u^i_0+\dot{v}_i+\underline{v}_i}\,\ud x&\ge&|\Omega|\exp\left(\frac{1}{|\Omega|}\ito\left(u^i_0+\dot{v}_i +\underline{v}_i\right)\ud x\right)\nonumber\\
 &=&|\Omega|\exp\left(\frac{1}{|\Omega|}\ito u^i_0\ud x\right)\re^{\underline{v}_i}\equiv\sigma_i\re^{\underline{v}_i}, \quad  i=1,\dots,m.\label{c17}
 \eer

Using  the condition \eqref{a10}, we have $K_i>0,i=1, \dots, m$.  Then, combining \eqref{c16} and \eqref{c17}, we have
   \ber
   I(\mathbf{w})-\frac{1}{2\lambda}\sum\limits_{i=1}^m\ito|\nabla \dot{w}_i|^2\ud x
   &\ge& \sum\limits_{i=1}^m(\sigma_i\re^{\underline{v}_i}-K_i\underline{v}_i) \nonumber\\
   &\ge& \sum\limits_{i=1}^mK_i\ln\frac{\sigma_i}{K_i}.\label{c18}
  \eer
Hence,  from \eqref{c18}, we see that the functional $I$ is bounded  from below and the minimization problem
 \ber
 \eta_0\equiv \inf\big\{ I(\mathbf{w}) | \mathbf{w}\in W^{1, 2}(\Omega)\big\} \label{c19}
 \eer
is well-defined.

Let $\{\mathbf{w}^{(\ell)}\}=\Big\{\Big(w_1^{(\ell)},\dots,w_m^{(\ell)}\Big)\Big\}$ be a minimizing sequence of \eqref{c19}.  It is easy to see  that the
function $f(t)=\sigma e^t-\eta t$, where $\sigma, \eta $ are positive constants, satisfies the property that  $f(t)\to \infty$ as $t\to \pm\infty$.
Thus, we  conclude  from \eqref{c18} that $\Big\{\underline{v}_i^{(\ell)}\Big\}\, ( i=1,\dots, m)$ are bounded. As a result, $\Big\{\underline{w}_i^{(\ell)}\Big\} \,(i=1,\dots, m)$ are bounded.
Therefore, the sequences $\Big\{\underline{w}_i^{(\ell)}\Big\} \,(i=1,\dots, m)$ admit convergent  subsequences, which are still denoted by $\Big\{\underline{w}_i^{(\ell)}\Big\}$ $(i=1,\dots, m)$ for convenience. Then,
there exist $m$  real numbers $\underline{w}_1^{(\infty)},\dots, \underline{w}_m^{(\infty)}\in \mathbb{R}$ such that $\underline{w}_i^{(\ell)}\to\underline{w}_i^{(\infty)} (i=1, \dots, m)$,  as $\ell\to \infty$.

Using \eqref{c18} again, we  infer that $\Big\{\nabla \dot{w}_i^{(\ell)}\Big\}\, (i=1,\dots, m)$  are bounded in $L^2(\Omega)$.
Therefore, it follows from the  Poincar\'{e} inequality that the sequences $\Big\{\dot{w}_i^{(\ell)}\Big\} \,(i=1,\dots, m)$ are bounded
in $W^{1, 2}(\Omega)$.  Consequently,  the sequences $\Big\{\dot{w}_i^{(\ell)}\Big\} \,( j=1,\dots, m)$ admit weakly convergent subsequences, which are
still denoted by $\Big\{\dot{w}_i^{(\ell)}\Big\}\, (i=1,\dots, m)$ for convenience. Then, there exist $m$ functions $\dot{w}_i^{(\infty)} \in W^{1,2}(\Omega)\, (i=1,\dots, m)$ such that $\dot{w}_i^{(\ell)}\to\dot{w}_i^{(\infty)}$  weakly in
$W^{1,2}(\Omega)$ as $\ell\to\infty \,(i=1,\dots, m)$. Of course, $\dot{w}_i^{(\infty)}\in \dot{W}^{1, 2}(\Omega)\, (i=1,\dots, m)$.

Set $w_i^{(\infty)}=\underline{w}_i^{(\infty)}+\dot{w}_i^{(\infty)}\, (i=1,\dots, m)$, which are all in $W^{1,2}(\Omega)$ naturally. Then, the above
convergence result implies $w_i^{(\ell)}\to w_i^{(\infty)}\, (i=1,\dots, m)$ weakly in $W^{1,2}(\Omega)$ as $\ell\to \infty$.
Noting that  the functional $I(\mathbf{w})$ is weakly lower semi-continuous in $W^{1,2}(\Omega)$,  we conclude that $\big(w_1^{(\infty)},\dots, w_m^{(\infty)}\big)$ is a
solution of the minimization problem \eqref{c19} and is a critical point of $I(\mathbf{w})$.  As a critical point of $I(\mathbf{w})$, it satisfies the equations \eqref{c9}--\eqref{c11}.

Since  the matrix  $R$ is positive definite, it is easy to check  that $I(\mathbf{w})$ is strictly convex  over  $W^{1,2}(\Omega)$. As a result, the functional
$I(\mathbf{w})$ has at most one critical point in $W^{1,2}(\Omega)$, which implies the uniqueness of the solution to the equations \eqref{c9}--\eqref{c11}.

\medskip 

As in \cite{liey1}, we briefly remark that we can also find a critical  point of the  functional $I$ by a constrained minimization procedure.

To proceed, we rewrite the constraints \eqref{c7} as
  \ber
    J_i (\mathbf{w})\equiv\ito \re^{u_0^i+v_i}\,\ud x= K_i, \quad i=1,\dots, m. \label{c20'}
  \eer
By the assumption \eqref{a10}, we have $K_i>0, i=1,\dots,   m$. Then we  consider the constrained minimization problem
  \ber
   \eta_0\equiv \inf\big\{ I(\mathbf{w}) |\, \mathbf{w}\in W^{1, 2}(\Omega) \,\,\text{and satisfies  }\, \eqref{c20'} \big\}. \label{c20}
  \eer
 If the problem \eqref{c20} has a  solution, say, $\tilde{\mathbf{w}}=(\tilde{w}_1, \dots, \tilde{w}_m)$, then there exist  some numbers (the Lagrangian multipliers)
 $\mu_1, \dots, \mu_m\in \mathbb{R}$  such that
 \ber
  \big(D(I+\mu_1J_1+\mu_2J_2+\dots+\mu_mJ_m)(\tilde{\mathbf{w}})\big)(\mathbf{w})=0, \quad
  \forall\,\mathbf{w}\,\in W^{1,2}(\Omega). \label{c21}
 \eer
 Using a series of test configurations $\mathbf{w}^i=(\delta_{i1}, \dots, \delta_{im}),  i=1, \dots,  m$ in \eqref{c21} successively, we have
  \be
  \left. \begin{array}{rrl} L_{11}\mu_1+L_{21}\mu_2&=&0,\\
   L_{ii}\mu_i+L_{i+1i}\mu_{i+1}&=&0,\quad i=2, \dots, m-1,\\
   L_{mm}\mu_m&=&0,\end{array}\right\}
  \ee
which imply
$\mu_1=\mu_2=\dots=\mu_m=0.$
This is to say that,  the constraints do not lead to  the undesired Lagrangian multiplier problem, and any solution of the constrained
minimization problem  \eqref{c21} is a critical point of the functional \eqref{c14} itself.

From the constraint  \eqref{c20'}, we see that
 \ber
  &&\underline{v}_1= L_{11}\underline{w}_1=\ln K_1-\ln\ito\re^{u_0^1+L_{11}\dot{w}_1}\ud x, \label{c21'}\\
  && \underline{v}_i= L_{ii-1}\underline{w}_{i-1}+ L_{ii}\underline{w}_{i}=\ln K_i-\ln\ito\re^{u_0^i+L_{ii-1}\dot{w}_{i-1}+L_{ii}\dot{w}_i}\ud x, \quad i=2, \dots, m. \label{c22}
  \eer
From  \eqref{c16}, we have
  \ber
  I(\mathbf{w})-\frac{1}{2\lambda}\sum\limits_{i=1}^m\|\nabla\dot{w}_i\|_2^2=\sum\limits_{i=1}^m\ito\re^{\underline{v}_i}\re^{u_0^i+\dot{v}_i}\ud  x
  -\sum\limits_{i=1}^m\ito K_i\underline{v}_i\ud x,\label{c23}
  \eer
Plugging \eqref{c22} into  \eqref{c23}, and using Jensen's inequality, we have
  \ber
  I(\mathbf{w})-\frac{1}{2\lambda}\sum\limits_{i=1}^m\|\nabla\dot{w}_i\|_2^2\ge-\sum\limits_{i=1}^m K_i \ln K_i+\sum\limits_{i=1}^m K_i\ln\sigma_i=
     \sum\limits_{i=1}^m K_i\ln\frac{\sigma_i}{K_i}, \label{c24}
  \eer
 where $\sigma_i=|\Omega|\exp\left(\frac{1}{|\Omega|}\ito u_0^i\ud x\right), i=1, \dots, m$.
 From \eqref{c24} we know that the functional $I$  is bounded from below.

 Let $\{\mathbf{w}^{(\ell)}\}=\Big\{\Big(w_1^{(\ell)},\dots, w_m^{(\ell)}\Big)\Big\}$ be a minimizing sequence of the problem \eqref{c20}. We conclude from \eqref{c24} that
 $\Big\{\dot{w}_i^{(\ell)}\Big\} \, (i=1,\dots m)$  are bounded in $W^{1, 2}(\Omega)$. Without loss of generality, we may assume  $\Big\{\dot{w}^{(\ell)}_i\Big\} \, (i=1,\dots m)$
 converge weakly in $W^{1, 2}(\Omega)$. The Trudinger--Moser inequality \eqref{c8} and \eqref{c21'} and \eqref{c22}  imply that $\Big\{\underline{w}^{(\ell)}_i\Big\} \, (i=1,\dots m)$   also converge.
 Therefore, the sequence  $\Big\{\Big(w_1^{(\ell)}, \dots, w_m^{(\ell)}\Big)\Big\}$  has a weak limit 
in $W^{1,2}(\Omega)$ as $\ell\to\infty$.  Noting that the constraint
 functionals are weakly continuous and the functional $I$ is weakly lower semi-continuous, we see that the weak limit of $\Big\{\Big(w_1^{(\ell)}, \dots, w_m^{(\ell)}\Big)\Big\}$
 must be a solution of the problem \eqref{c20}. Since $I$ is strictly convex, this solution is unique. Thus the constrained minimization procedure is carried out as well.

\medskip 

Finally, integrating the equations \eqref{c1}--\eqref{c3} or \eqref{c4} over $\Omega$, we see that the
  quantized integrals \eqref{a9} are established.

\section{The special case where $a=0$}
\setcounter{equation}{0} \setcounter{theorem}{0}

When $a=0$, we denote the matrix  $R$ by  $\dot{R}$. Then
 \begin{equation}
\dot{R}=\begin{pmatrix}
 1 & -1 & 0&\dots&\dots & 0\\
-1 &  3 & -2&0&\dots & 0\\
0 &-2&  5 &-3&\dots&0\\
\vdots& &\ddots&\ddots&\ddots&\vdots\\
0&\quad &\ddots&-(m-2)&2m-3&-(m-1)\\
 0 & \dots &   &0&-(m-1)&2m-1
\end{pmatrix}.
\end{equation}

By a direct calculation,  we see that the  leading principal minors of $\dot{R}$ are
  \be 
\dot{R}_i=i!,\quad i=1,\dots, m.\ee
Then by our formula  \eqref{b4} for   the inverse of $\dot{R}$, we  have
\be 
(\dot{R}^{-1})_{ij}=\sum\limits_{l=j}^m\frac1l, \quad j=i, i+1, \dots,m,\quad i=1,\dots,m. \ee
 that is
 \begin{equation}
\dot{R}^{-1}=\begin{pmatrix}
 \sum\limits_{l=1}^m\frac1l& \sum\limits_{l=2}^m\frac1l & \sum\limits_{l=3}^m\frac1l&\dots& \frac1m\\
 \sum\limits_{l=2}^m\frac1l &\sum\limits_{l=2}^m\frac1l  & \sum\limits_{l=3}^m\frac1l &\dots& \frac1m\\
 \sum\limits_{l=3}^m\frac1l &\sum\limits_{l=3}^m\frac1l&\sum\limits_{l=3}^m\frac1l &\dots& \frac1m\\
 \vdots& \vdots&\vdots&\ddots &\vdots \\
 \frac1m &\frac1m  & \frac1m  &\dots& \frac1m
\end{pmatrix}.
\end{equation}

From this expression of $\dot{R}^{-1}$, we have
 \be 
r_i\equiv\sum\limits_{j=1}^m(\dot{R}^{-1})_{ij}=m-i+1, \quad i=1,\dots, m.
\ee

Thus,  applying  Theorem \ref{th1}, we obtain Theorem \ref{th2}.
\medskip 

An interesting feature of the case $a=0$ is that the matrix-valued charge densities $j^0$ and $J_{12}^0$
may coincide to satisfy the relation \cite{kkkn}
\be 
j^0=J_{12}^0=\frac k{2\pi}B,
\ee
so that, using (\ref{flux}), the  minimum energy is seen to be directly related to the vortex numbers according to the quantization formula
\ber 
E&=&\frac12\mu|Q+2R_{12}|=\frac13\mu\left|\mbox{Tr}\int j^0\,\ud x+2\mbox{Tr}\int J_{12}^0\,\ud x\right|\nm\\
&=&\frac{k\mu}{2\pi}\left|\mbox{Tr}\int B\,\ud x\right|=k\mu\sum_{i=1}^{N-1}(N-i)n_i,
\eer
as stated in \cite{kkkn}.

\section{Some concrete examples}
\setcounter{equation}{0}

When $a>0$, the matrix computation quickly becomes rather involved for large $N$ and the results are not as explicit as the case
for $a=0$.
However, if $N$ is low and concrete, we can readily apply Theorem \ref{th1} to obtain
explicit results for the problem. As an illustration,  we work out the equations for  $N=3$ (or $m=2$) and $N=4$
(or $m=3$), respectively, as examples.

We first consider the case when $m=2$. The equations in this case are
 \ber
  \Delta u_1&=&\lambda\left([2a^2+1]\re^{u_1}-[a^2+1]\re^{u_2}-1\right) +4\pi\sum\limits_{s=1}^{n_1}\delta_{p_{1,s}},\label{z1}\\
   \Delta u_2&=&\lambda\left(-[a^2+1]\re^{u_1}+[2a^2+3]\re^{u_2}-1\right) +4\pi\sum\limits_{s=1}^{n_2}\delta_{p_{2,s}}, \label{z2}
 \eer
for which the associated coefficient matrix $R$ reads
  \begin{equation}
 R=\begin{pmatrix}
 2a^2+1 & -(a^2+1) \\
 -(a^2+1)&  2a^2+3
\end{pmatrix},
\end{equation}
with the eigenvalues
   \be
     \lambda_{1,2}= 2a^2\pm\sqrt{a^4+2a^2+2}+2.
    \ee
Therefore
    \be
    \lambda_0 =2\left( 2a^2-\sqrt{a^4+2a^2+2}+2\right).\label{z2'}
    \ee

On the other hand, the inverse of $R$ is
 \be
 R^{-1}=\frac{1}{3a^4+6a^2+2}\begin{pmatrix}
 2a^2+3 & a^2+1 \\
 a^2+1&  2a^2+1
 \end{pmatrix}.  \label{z3}
\ee
Thus, by definition, we have
  \ber
   r_1=\frac{3a^2+4}{3a^4+6a^2+2}, \quad     r_2=\frac{3a^2+2}{3a^4+6a^2+2}.\label{z4}
  \eer

In view of Theorem \ref{th1}, we obtain the following explicit results.

\begin{theorem}\label{th51}
  For any $a\ge0, \lambda>0$, consider  the   equations \eqref{z1}--\eqref{z2}.

 (i) The  equations \eqref{z1}--\eqref{z2} have a unique solution over $\mathbb{R}^2$  satisfying the boundary condition
    \be
     u_1\to \ln\frac{3a^2+4}{3a^4+6a^2+2},  \quad u_2\to \ln\frac{3a^2+2}{3a^4+6a^2+2},\quad |x|\to \infty. \label{z5}
    \ee
 Moreover, this solution obeys the following exponential decay estimate near infinity:
    \ber
     \sum\limits_{i=1}^2(u_i(x)-\ln r_i)^2\le
     C(\vep)\re^{-(1-\vep)\sqrt{\lambda\lambda_0}|x|},\label{z6}
      \eer
 where $\vep\in(0,1)$ is arbitrarily small, $C(\vep)$ is a positive constant depending on $\vep$, 
$r_1,r_2$ are given by \eqref{z4}, and $\lambda_0$ is given by \eqref{z2'}.

 (ii) The equations \eqref{z1}--\eqref{z2} over a doubly periodic domain $\Omega$ have a 
 solution if and only if the following two inequalities
 \ber
    4\pi\left([2a^2+3]n_1+[a^2+1]n_2\right)&<&\lambda|\Omega|(3a^2+4), \label{z7}\\
    4\pi\left([a^2+1]n_1+[2a^2+1]n_2\right)&<&\lambda|\Omega|(3a^2+2), \label{z8}
 \eer
hold simultaneously. Furthermore, if a solution exists, it must be unique.

\end{theorem}

Next we consider the case when $m=3$. The equations are
 \ber
  \Delta u_1&=&\lambda\left([2a^2+1]\re^{u_1}-[a^2+1]\re^{u_2}-1\right) +4\pi\sum\limits_{s=1}^{n_1}\delta_{p_{1,s}},\label{z9}\\
   \Delta u_2&=&\lambda\left(-[a^2+1]\re^{u_1}+[2a^2+3]\re^{u_2}-[a^2+2]\re^{u_3}-1\right) +4\pi\sum\limits_{s=1}^{n_2}\delta_{p_{2,s}}. \label{z10}\\
     \Delta u_3&=&\lambda\left(-[a^2+2]\re^{u_2}+[2a^2+5]\re^{u_3}-1\right) +4\pi\sum\limits_{s=1}^{n_3}\delta_{p_{3,s}}, \label{z11}
 \eer
so that the associated coefficient matrix $R$ takes the form
  \begin{equation}
 R=\begin{pmatrix}
 2a^2+1 & -(a^2+1)&0 \\
 -(a^2+1)&  2a^2+3&-(a^2+2)\\
 0&-(a^2+2)& 2a^2+5
\end{pmatrix}.
\end{equation}
We will not compute all the eigenvalues of $R$. But, instead, we write down
the inverse of $R$:
 \be
 R^{-1}=\frac{1}{4a^6+18a^4+22a^2+6}\begin{pmatrix}
  3a^4+12a^2+11 & 2a^4+7a^2+5&   a^4+3a^2+2 \\
  2a^4+7a^2+5 &  4a^4+12a^2+5&  2a^4+5a^2+2\\
  a^4+3a^2+2 &  2a^4+5a^2+2&  3a^4+6a^2+2
 \end{pmatrix},  \label{z12}
\ee
which gives us 
  \ber
  r_1&=&\frac{3a^4+11a^2+9}{2a^6+9a^4+11a^2+3},   \label{z13} \\
  r_2&=&\frac{4a^4+12a^2+6}{2a^6+9a^4+11a^2+3},  \label{z13a}\\
   r_3&=&\frac{3a^4+7a^2+3}{2a^6+9a^4+11a^2+3}. \label{z13b}
  \eer

Applying Theorem \ref{th1} again, we can state

\begin{theorem}\label{th52}
  For any $a\ge0, \lambda>0$, consider  the  equations \eqref{z9}--\eqref{z11}.

 (i) The  equations   have a unique solution over $\mathbb{R}^2$  satisfying the boundary condition
    \be
     u_i\to \ln r_i,  \quad |x|\to \infty, \quad i=1,2,3,\label{z14}
    \ee
    where $r_1,r_2,r_3$ are defined by \eqref{z13}--\eqref{z13b}.
Moreover, this solution satisfies the following exponential decay estimate at infinity:
    \ber
     \sum\limits_{i=1}^3(u_i(x)-\ln r_i)^2\le
     C(\vep)\re^{-(1-\vep)\sqrt{\lambda\lambda_0}|x|},\label{z15}
      \eer
 where $\vep\in(0,1)$ is small, $C(\vep)>0$ depends on $\vep$,  and  $\lambda_0$ is twice the smallest eigenvalue of $R$.

 (ii) The equations \eqref{z9}--\eqref{z11}  over a doubly periodic domain $\Omega$
have a
 solution if and only if the following three inequalities
   \ber
    2\pi\left([3a^4+12a^2+11]n_1+[2a^4+7a^2+5]n_2+[a^4+3a^2+2]n_3\right)&<&\lambda|\Omega|(3a^4+11a^2+9),\quad \quad\label{z16}\\
     \pi\left([2a^4+7a^2+5]n_1+[4a^4+12a^2+5]n_2+[2a^4+5a^2+2]n_3\right)&<&\lambda|\Omega|(2a^4+6a^2+3), \quad\label{z17}\\
        2\pi\left([1a^4+3a^2+2]n_1+[2a^4+5a^2+2]n_2+[3a^4+6a^2+2]n_3\right)&<& \lambda|\Omega|(3a^4+7a^2+3),\quad \label{z18}
 \eer
hold simultaneously. Furthermore, if a solution exists, it must be unique.
\end{theorem}

In general, for any $N\geq2$, since we have established the existence and uniqueness of a solution realizing $n_i$ arbitrarily
prescribed zeros for each $f_i$, $i=1,\dots, N-1$, we see that the solutions precisely depend on
\be 
2n\equiv 2\sum_{i=1}^{N-1}n_i
\ee
continuous parameters which are the coordinates of the vortex points 
\be 
p_{1,1},\dots,p_{1,n_1},\dots,p_{N-1,1},\dots,p_{N-1,n_{N-1}}.
\ee

It may be relevant to recall the index-theory study of Weinberg \cite{We} on the Abelian Higgs BPS equations
which shows that the dimension of the moduli space of multiple vortex solutions is $2n$. This number is precisely
the number of the coordinates in $\mathbb{R}^2$ 
which are needed to determine the locations of $n$ vortices, in view of the classical existence and uniqueness results in \cite{jata}. There are similar studies in
the context of non-relativistic Chern--Simons equations \cite{KSY} and of superconducting cosmic strings \cite{Sem}. It
will be interesting to know the dimension of the moduli space of the multiple vortex solutions of
the BPS vortex equations (\ref{BPS})--(\ref{e21}) of Kim--Kim--Kwon--Nakajima \cite{kkkn} in the ABJM theory.
Thus our existence and uniqueness results
indicate that, within the ansatz of Kim--Kim--Kwon--Nakajima \cite{kkkn}, the dimension of the moduli space 
of the $n$-vortex solutions of the BPS equations (\ref{BPS})--(\ref{e21})  is also $2n$, although, generally
when no specific ansatz is specified, it should
depend on the gauge group index $N$ as well \cite{Eto}.


\begin{thebibliography}{99}

\bibitem{Ab}
A. A. Abrikosov, On the magnetic properties of superconductors of the second group,
{\em Sov. Phys. JETP} {\bf5} (1957) 1174--1182.

\bibitem{abjm}O. Aharony, O. Bergman, D. L. Jaferis and J. Maldacena, $\mathcal{N}=6$ superconformal
Chern-Simons-matter theories, M2-branes and their gravity duals, {\em J. High Energy Phys.} 0810 (2008)  091.


\bibitem{aubi}T. Aubin, {\em Nonlinear Analysis on Manifolds: Monge-Amp\'{e}re Equations}, Springer, Berlin and New York, 1982.

\bibitem{ak}R. Auzzi and S. P. Kumar,  Non-Abelian vortices at weak and strong coupling in mass deformed ABJM theory.  {\em J. High Energy Phys.}  {\bf 071}   0910  (2009).

\bibitem{BL1}
J. Bagger and N. Lambert, Modeling multiple M2's, {\em Phys. Rev.} D {\bf75} (2007) 045020.


\bibitem{BL2}
 J. Bagger and N. Lambert, Gauge symmetry and supersymmetry of multiple M2-branes, {\em Phys. Rev.} D {\bf77} (2008) 065008.

\bibitem{BL3}
J. Bagger and N. Lambert, Comments on multiple M2-branes, {\em J. High Energy Phys.} {\bf0802} (2008)
105.

\bibitem{BLS}
M. A. Bandres, A. E. Lipstein, and J. H. Schwarz,
Studies of the ABJM theory in a formulation with manifest $SU(4)$ $R$-symmetry,
{\em J. High Energy Phys.}  0809 (2008) 027. 

\bibitem{BBKR}
D. Belyaev, L. Brink, S.-S. Kim, and P. Ramond,
The BLG theory in light-cone superspace,
{\em J. High Energy Phys.} {\bf1004} (2010) 026.

\bibitem{Bo}
E. B. Bogomol'nyi, The stability of classical solutions, {\em Sov. J. Nucl. Phys.} {\bf24} (1976) 449--454.

\bibitem{CY}
L. Caffarelli and Y. Yang, Vortex condensation in the Chern--Simons Higgs model: an existence
theorem, {\em Commun. Math. Phys.} {\bf168} (1995) 321--336.

\bibitem{Chae}
D. Chae and O. Yu. Imanuvilov, 
The existence of nontopological multivortex solutions
in the relativistic self-dual Chern--Simons theory, 
{\em Commun. Math. Phys.} {\bf215} (2000)  119--142.

\bibitem{CCR}
 S. Chakrabortty, S. P. Chowdhury, and K. Ray,
Some BPS configurations of the BLG theory,
{\em Phys. Lett.} B {\bf703} (2011) 172--179. 

\bibitem{CFL}
H. Chan, C. C. Fu, and C. S. Lin, Non-topological multivortex solutions to the self-dual Chern--Simons--Higgs
equation, {\em Commun. Math. Phys.} {\bf231} (2002) 189--221.

 
\bibitem{chya1}S. Chen and Y. Yang, Existence of multiple vortices in supersymmetric gauge field theory, {\it Proc. R. Soc.} A, in press.

\bibitem{Dunne}
G. Dunne, {\em Self-Dual Chern--Simons Theories}, Lecture Notes in Physics, vol. m {\bf 36},
Springer, Berlin, 1995.

\bibitem{Eto}
M. Eto, Y. Isozumi, M. Nitta, K. Ohashi, and N. Sakai,
Moduli space of non-Abelian vortices,
{\em  Phys. Rev. Lett.} {\bf96} (2006) 161601.

\bibitem{EMP}
B. Ezhuthachan, S. Mukhi, and C. Papageorgakis,
The power of the Higgs mechanism: higher-derivative BLG theories,
{\em J. High Energy Phys.} {\bf0904} (2009) 101. 

\bibitem{font}L. Fontana, Sharp borderline Sobolev inequalities on compact Riemannian manifolds, {\em Comment. Math. Helv.} {\bf68} (1993) 415--454.

\bibitem{F}
J. Fr\"{o}hlich, The fractional quantum Hall effect, Chern--Simons theory, and integral
lattices, {\em Proc. Internat. Congr. Math.}, pp. 75--105, Birkh\"{a}user, Basel, 1995.

\bibitem{FM1}
J. Fr\"{o}hlich and P. Marchetti, Quantum field theory of  anyons, {\em
Lett. Math. Phys.} {\bf16} (1988) 347--358.

\bibitem{FM2}
J. Fr\"{o}hlich and P. Marchetti, Quantum field theory of vortices and anyons, {\em
Commun. Math. Phys.} {\bf121} (1989) 177--223.




\bibitem{goor}G. H. Golub and J. M. Ortega, {\em Scientific Computing and Differential Equations}, Academic, San Diego, 1992.

\bibitem{GST}
A. Gustafson, I. M. Sigal, and T. Tzaneteas, Statics and dynamics of magnetic vortices and of Nielsen--Olesen (Nambu)
strings, {\em J. Math. Phys.} {\bf51} (2010) 015217.

\bibitem{G}
A. Gustavsson, Algebraic structures on parallel M2-branes, {\em Nucl. Phys.} B {\bf811}
(2009) 66--76.

\bibitem{hoof}
G. 't Hooft, A property of electric and magnetic flux in non-Abelian gauge theories, {\em Nucl. Phys. B} {\bf153} (1979) 141--160.

\bibitem{hll} K. Hosomichi,  K. Lee, and S. Lee, Mass-deformed Bagger--Lambert theory and its BPS objects, Phys. Rev. D {\bf78} (2008) 066015.

\bibitem{HKP}
J. Hong, Y. Kim and P.-Y. Pac, Multivortex solutions of the Abelian Chern--Simons--Higgs
theory, {\em Phys. Rev. Lett.} {\bf64} (1990) 2330--2333.

\bibitem{JPW}
R. Jackiw, S.-Y. Pi, and E. J. Weinberg, Topological and non-topological solitons in relativistic
and non-relativistic Chern--Simons theory, {\em Particles, Strings and Cosmology} (Boston, 1990),
pp. 573--588, World Sci. Pub., River Edge, NJ, 1991.


\bibitem{JW}
R. Jackiw and E. J. Weinberg, Self-dual Chern--Simons vortices, {\em Phys. Rev. Lett.}
{\bf64} (1990) 2334--2337.

\bibitem{jata} 
A. Jaffe and C. H. Taubes, {\em Vortices and Monopoles}, Birkh\"{a}user, Boston, 1980.

 \bibitem{kim1}C.  Kim, Vortex-type solutions in ABJM theory, {\em Journal of Physics:  Conference Series} {\bf 343} (2012) 012057.

\bibitem{kkkn} C. Kim, Y. Kim,  O. K. Kwon, and H. Nakajima, Vortex-type half-BPS solitons in ABJM  theory, {\em Phys. Rev. D} {\bf 80}  (2009)  045013.

\bibitem{KSY}
S. K. Kim, K. S. Soh, and J. H. Yee,
Index theory for the nonrelativistic Chern--Simons solitons
{\em Phys. Rev.} D {\bf42} (1990) 
4139--4144.

\bibitem{liey1}E. H. Lieb and Y. Yang, Non-Abelian vortices in supersymmetric gauge field theory via direct methods, {\em Commun. Math. Phys.}
 {\bf 313} (2012) 445--478

\bibitem{liny1}C. S.  Lin and Y. Yang,  Non-Abelian multiple vortices in supersymmetric field theory, {\em Comm. Math. Phys.} {\bf 304} (2011) 433--457,

\bibitem{liny2} C. S. Lin and Y. Yang, Sharp existence and uniqueness theorems for non-Abelian multiple vortex solutions, {\em  Nucl. Phys.} B {\bf 846} (2011) 650--676.

\bibitem{NO}
H. Nielsen and P. Olesen, Vortex-line models for dual strings, {\em Nucl. Phys.} B
{\bf61} (1973)  45--61.

\bibitem{NT}
 M. Nolasco and G. Tarantello,  Vortex condensates for the $SU(3)$ Chern--Simons theory,
{\em  Commun. Math. Phys.} {\bf213} (2000) 599--639.

\bibitem{PS}
M. K. Prasad and C. M. Sommerfield, Exact classical solutions for the 't Hooft monopole and the Julia--Zee dyon,
{\em Phys. Rev. Lett.} {\bf35} (1975) 760--762.

\bibitem{Sch}
J. H. Schwarz,
Superconformal Chern--Simons theories,
{\em J. High Energy Phys.} {\bf 0411} (2004) 078.

\bibitem{Sem}
G. W.  Semenoff, Index theorems and superconducting cosmic strings,
{\em Phys. Rev.} D {\bf37} (1988) 2838--2852.

\bibitem{SYcs1}
J. Spruck and Y. Yang, Topological solutions in the self-dual Chern--Simons theory: existence
and approximation, {\em Ann. Inst. H. Poincar\'{e} -- Anal. non Lin\'{e}aire}
{\bf12} (1995) 75--97.

\bibitem{SYcs2}
J. Spruck and Y. Yang, The existence of non-topological solitons in the self-dual Chern--Simons
theory, {\em Commun. Math. Phys.} {\bf149} (1992)  361--376.

\bibitem{Ta}
G. Tarantello, Multiple condensate solutions for the Chern--Simons--Higgs theory,
{\em J. Math. Phys.} {\bf37} (1996)  3769--3796.

\bibitem{Tbook}
G. Tarantello, {\em Self-Dual Gauge Field Vortices}, Progress in Nonlinear Differential Equations and Their Applications
{\bf72}, Birkh\"{a}user, Boston, 2008.

\bibitem{T1}
C. H. Taubes, Arbitrary $N$-vortex solutions to the first order Ginzburg--Landau equations, {\em Commun. Math.
Phys.} {\bf72} (1980)  277--292.

\bibitem{waya}S. Wang and Y. Yang, Abrikosov's vortices in the critical coupling, {\em SIAM J. Math. Anal.} {\bf 23} (1992) 1125--1140.

\bibitem{We}
E. J. Weinberg, Multivortex solutions of the Ginzburg--Landau equations, 
{\em Phys. Rev.} D {\bf19} (1979)  3008--3012.

\bibitem{Wil}
F. Wilczek, {\em Fractional Statistics and Anyonic Superconductivity}, World Scientific, Singapore,
1990.

\bibitem{Ycs}
Y. Yang, The relativistic non-Abelian Chern--Simons equations, {\em Commun. Math. Phys.}
{\bf186} (1997)  199--218.

\bibitem{yang2}
Y. Yang, On a system of nonlinear elliptic equations arising in theoretical physics, {\em J. Funct.
Anal.} {\bf 170} (2000) 1--36.


\bibitem{yang1}
Y. Yang, {\em Solitons in Field Theory and Nonlinear Analysis}, Springer, New York, 2001.


\end{thebibliography}
\end{document}